\documentclass{article}

\usepackage{arxiv}

\usepackage[utf8]{inputenc}
\usepackage[T1]{fontenc}   
\usepackage{url,doi,hyperref}          
\usepackage{booktabs}       
\usepackage{amsfonts}       
\usepackage{nicefrac}       
\usepackage{microtype}      
\usepackage{graphicx}
\usepackage[numbers,square]{natbib}
\usepackage{amsfonts,amsmath,amssymb}
\usepackage[caption = false]{subfig}
\usepackage[algo2e,ruled,vlined]{algorithm2e}
\newcommand{\orcid}[1]{\href{https://orcid.org/#1}{\includegraphics[width=8pt]{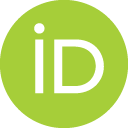}}}

\title{Introducing a Novel Data over Voice Technique \\ for Secure Voice Communication}

\author{%
  \orcid{0000-0002-4564-7946} Piotr Krasnowski,  \orcid{0000-0002-9590-164X} Jerome Lebrun, \orcid{0000-0002-0048-5197} Bruno Martin\\
  University C\^ote d'Azur, I3S-CNRS, 2000, route des Lucioles, \\
  06900 Sophia Antipolis, France \\
  \texttt{p.g.krasnowski@gmail.com}, \texttt{\{lebrun,bruno.martin\}@i3s.unice.fr} \\
}

\date{}

\renewcommand{\shorttitle}{Introducing a Novel Data over Voice Technique for Secure Voice Communication}

\begin{document}
\maketitle

\begin{abstract}
The increasing need for privacy-preserving voice communications is encouraging the investigation of new secure voice transmission techniques. This paper refers to the original concept of sending encrypted data or speech as pseudo-speech in the audio domain over existing voice communication infrastructures, like 3G cellular network and Voice over IP (VoIP). The distinctive characteristic of such a communication system is that it relies on the robust transmission of binary information in the form of audio signal.

This work presents a novel Data over Voice (DoV) technique based on codebooks of short harmonic waveforms. The technique provides a sufficiently fast and reliable data rate over cellular networks and many VoIP applications. The new method relies on general principles of Linear Predictive Coding for voice compression (LPC voice coding) and is more versatile compared to solutions trained on exact channel models. The technique gives by design a high control over the desired rate of transmission and provides robustness to channel distortion. In addition, an efficient codebook design approach inspired by quaternary error correcting codes is proposed.

The usability of the proposed DoV technique for secure voice communication over cellular networks and VoIP has been successfully validated by empirical experiments. The paper details the system parameters, putting a special emphasis on system's security and technical challenges.
\end{abstract}

\keywords{Secure voice communications \and Data over voice \and Digital voice channels \and VoIP \and Quaternary ECC}

\section{Introduction}
\label{intro}

The growing risk of privacy violation associated with the rapid spread of mobile communications has motivated the development of secure VoIP communicators, with Telegram and Signal being the iconic examples.\footnote{\url{https://core.telegram.org}, \url{https://signal.org}} However, these applications cannot protect against spying malware, which could be installed directly on smartphones \cite{scott2017reckless}. To prevent this risk, voice can be encrypted in the audio domain by an external unit, secured against remote interception. In such a setting, voice signal is firstly transformed into an encrypted data stream shaped as pseudo-speech. Then, the encrypted audio signal is sent over the network instead of true speech signal.

Parallely, cellular vocal networks, thanks to their high reliability and wide coverage, attracted attention as a potential high-priority, low-bandwidth data communication channel with errors. The work on Data over Voice (DoV) technology enabled new applications, such as emergency call system eCall \cite{werner2009cellular}, messaging over voice \cite{dhananjay2010hermes}, point of sell (POS) financial transactions \cite{mezgec2009implementation}, and secure data and voice communications~\cite{chen2011ofdm,katugampala2003secure}.

With the quickly expanding data-driven 5G networks, the use of voice channels for sending data diminishes. Nevertheless, DoV techniques are still crucial in secure voice communications, for example, provided by Crypto Phones or other specialized devices~\citep{krasnowski2020introducing}. On the other hand, voice channels can be maliciously used for extruding private data or in Advanced Persistent Threat (APT) attacks~\citep{lee2017vulnerability}. 

The crucial challenges related to DoV are a consequence of principles underlying digital voice channels. Namely, voice channels aim at preserving speech intelligibility and quality while reducing the perceptually redundant information. In contrast to classical data channels, voice channels significantly distort the sent signal due to transcodings and audio processing. Moreover, modern digital voice channels are selective to signal parameters conforming to the speech model adopted in a particular system. To mitigate signal degradation caused by voice channels, several authors proposed DoV techniques based on encoding the data signal into speech-like parameters, codebook training, or modulation techniques. 

Katugampala et al. \citep{katugampala2003secure} proposed a system that uses predefined codebooks to map bits into vocal parameters: energy, pitch, and spectral envelope (encoded as line spectral pairs, LSP \citep{soong1984lsp}). The encoded parameters are transformed into a pseudo-speech signal adapted to transmission over a cellular network. Data extraction is done by a paired speech analyzer, which restores vocal parameters from the signal and decodes codebook indices. The system enabled transmission over a real GSM voice channel at the rate of 3000 bps with 2.9\% BER  \citep{katugampala2005secure}. Similar techniques were presented by Ozkan et al. \citep{ozkan2015data}, and Rashidi et al. \citep{Rashidi_dov}, who achieved respectively transmission rates of 1600 bps and 2000 bps by simulations.   

LaDue et al. \citep{LaDue2008}, and Sapozhnykov and Fienberg \citep{sapozhnykov2012low} investigated genetic and pattern matching algorithms to construct codebooks of short speech-like waveforms. Instead of synthesizing pseudo-speech, the authors proposed encoding bitstream directly into a sequence of symbols selected from a trained wavetable. Upon reception, received symbols were decoded with a bank of matched filters. The technique achieved the bandwidth of 4000 bps with 2.3 \% BER over enhanced full rate (EFR) voice channel. Unfortunately, the training process was time-consuming and required considerable computational resources. Moreover, the obtained wavetable was compatible with a unique channel model and hence impractical in real communication.

The problem of long and heavy computations has been tackled by Shahbazi et al. \citep{shahbazi2009novel}, and Boloursaz et al. \citep{Boloursaz2013}, who simplified the codebook construction by limiting the search to signals from the TIMIT speech database \citep{Timit}. Parallelly, Kazemi et al. \citep{kazemi2015modem} proposed a new idea to exploit sphere packing techniques to construct waveforms with a large minimum distance and an improved detection rate. Very recently, Zhang et al. \citep{Zhang_DoV} showed an analogous DoV technique based on sphere surface packing.

Finally, there exists a range of DoV techniques based on well-established, classical signal modulation. Zhan Xu \citep{xu2017data}, Chmayssani and Baudoin \citep{chmayssani2008data} tested by simulations phase shift keying modulation (PSK) and quadrature amplitude modulation (QAM), and achieved bitrates within the range 1 - 3 kbps. Ali et al. \citep{Baudoin2013} exploited M-ary frequency shift keying (M-FSK), whereas Dhananjay et al. \citep{dhananjay2010hermes} introduced a modified binary FSK (BFSK) tolerant to a small frequency deviation. Chen and Guo \citep{chen2011ofdm} reported a solution using orthogonal frequency division multiplexing (OFDM) modulation combined with PSK. 

An inspiring technique based on Amplitude Shift keying (ASK), named PCCD-OFDM-ASK, has been presented by Mezgec et al. \citep{mezgec2009implementation}. Phase-Continuity and Context Dependency (PCCD) refers to techniques providing phase continuity of the modulated signal. In PCCD-OFDM-ASK, blocks of 8-bit sequences are encoded onto eight orthogonal harmonics, numbered from 1 to 8. In contrast to classical OFDM, each bit in the 8-bit block is represented by the presence or absence of an orthogonal carrier. For instance, the binary 8-bit sequence `10001010' is mapped to a symbol with harmonics present only at positions 1, 5, and 7.  The scheme offers robust transmission up to 500 bps over real cellular voice channels.  

This article introduces a new DoV codebook-based modulation over cellular networks and VoIP for the needs of secure voice communication. The novelty comes from our simplified and universal codebook design process compared with the usual extensive codebook training on a selected voice model. Nevertheless, the method can be adapted to a particular channel, avoiding codebook over-tuning in the presence of fluctuating channel characteristics. Modulation parameters are easily adjustable in order to balance the transmission bitrate and the robustness to errors. 

The proposed technique was thoroughly tested with real voice calls. The scheme achieves up to 6.4 kbps over VoIP voice channels using 4G wireless network and 2.4 kbps over 3G cellular calls (see Section \ref{real_section}). It also enables safe voice transmission with an effective binary error rate significantly below 1\%.

This paper is organized as follows. Section \ref{section2} outlines challenges related to sending data over voice channels with LPC-based speech compression. Section \ref{section3} investigates signal distortion introduced by three selected LPC coders: AMR, Speex, and Opus-Silk. Next, the section describes the novel DoV technique, including codebook construction, signal generation, and demodulation. Section \ref{Section_experiments} presents performance results obtained by simulations and real-world experiments, and Section \ref{Section_voice} proposes a secure voice communication scheme using DoV. Finally, Section \ref{Section_conclusion} concludes the article and gives prospects for a future work.

\section{Digital Voice Channels}
\label{section2}

This section introduces crucial challenges related to data transmission over voice channels. It outlines the specific behavior of voice channels, very different compared to classical communication channels, and highlights the desired properties of DoV signals.

\subsection{Voice channel characteristics}

In real-world implementations, a complete voice channel is typically the concatenation of algorithms that transform a speech signal into binary data suitable for transmission over the network. Despite the lossy nature of speech processing, the received binary information is sufficient to re-synthesize a speech perceptually similar to the initial. However, from a DoV perspective, it is more convenient to consider voice channels as communication channels with particular constraints and signal distortion characteristics.

The core elements of any digital voice channel are \emph{voice coders}, which compress and encode sampled speech waveform exploiting principles of speech production and perception \citep{rabiner2011theory}. Real-time voice coders usually process speech on a frame basis by mapping portions of a speech waveform into sets of vocal parameters. These algorithms may perform high-pass filtering, differential encoding, and adaptive quantization to improve the compression ratio depending on the available network throughput. Unfortunately, such operations add memory and latency to a voice channel, and make it non-linear and non-stationary. 

In addition to voice compression, modern voice communication systems apply techniques such as \emph{voice activity detection} (VAD) \citep{backstrom2017speech}, \emph{adaptive gain control} (AGC) \citep{heitkamper1995optimization} or \emph{noise suppression} (NS) \citep{tsoukalas1997speech}. In opposition to voice coders, the implementation of these algorithms is rarely public and their impact on the DoV cannot be fully predicted.

Combining all the mentioned elements of real voice channels, achieving an analytic model of signal distortion is usually intractable. Nevertheless, it is still worthwhile to consider the most fundamental properties of voice channels and construct the DoV scheme agnostic to small variations of the voice channel characteristics.

\subsection{LPC coders}

Most of the voice coders operating in the upper-middle bitrate range (10 kbps \textendash 16 kbps) listed in ITU, IETF and 3GPP standards, and which are widely adopted in cellular and VoIP systems, rely on \emph{linear predictive coding} (LPC). LPC coders take their inspiration from the simplified speech production model, often referred to as a source-filter model \citep{AcousticTheoryofSpeechProduction,lochbaum1962speech}. According to the model, voice sound originates from a single source $e(t)$ and is filtered by a vocal tract with an impulse response $v(t)$. Such a simplification is justified for voiced and stationary sounds, which can be approximately represented by the buzzing excitation produced in the glottis and shaped when passing through the pharynx and between tongue, teeth, and lips. The resulting signal has the form $s(t)=e(t)*v(t)$, where $*$ denotes the convolution product.
 
Furthermore, considering voice as the convolution of excitation and vocal tract shaping would be of little practical value without effective methods for separating these components. The excitation and vocal tract characteristics can be well approximated during LPC analysis (hence LPC coders). The outputs of LPC analysis consist of a linear prediction filter describing the vocal tract's filtering effect and a residual that can be viewed as an excitation signal.

  \begin{figure}[h]
 	\centering
 	\subfloat[]{\label{fig:one1}
 	   {\includegraphics[width=.45\textwidth]{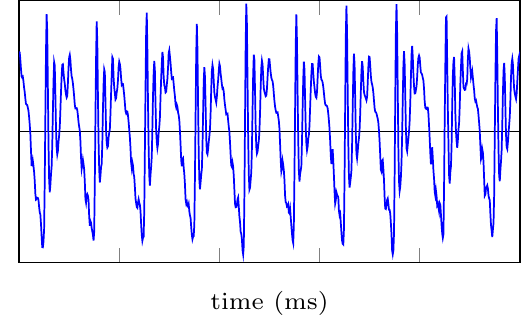}}
 	}
 	\hfill
 	\subfloat[]{\label{fig:two1}
 	{\includegraphics[width=.45\textwidth]{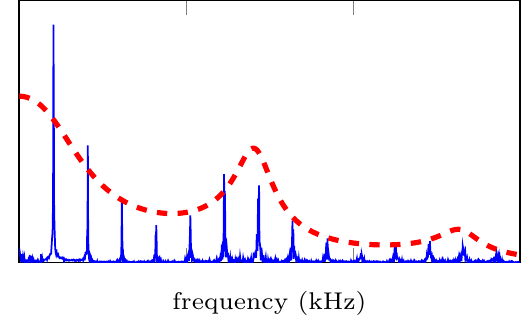}}
 	}
 	\\
 	\subfloat[]{\label{fig:three1}
 	{\includegraphics[width=.45\textwidth]{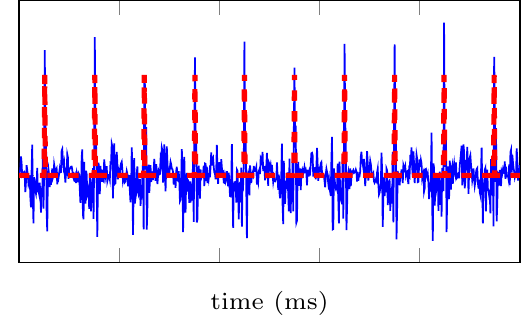}}
 	}
 	\hfill
 	\subfloat[]{\label{fig:four1}
 	{\includegraphics[width=.45\textwidth]{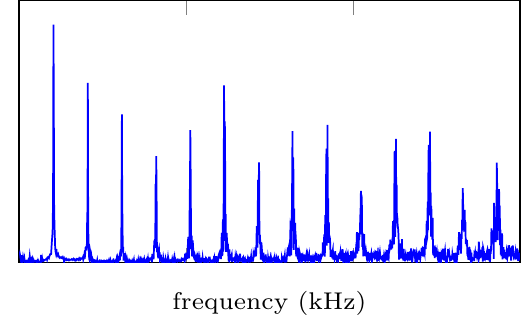}}
 	}
 	\caption{LPC analysis of vowel $/a/$: (a) time domain waveform, (b) spectrum of the waveform (blue solid line) and frequency response of the 12th order LPC filter (red dashed line), (c) residual of LPC analysis (solid blue line) and excitation peaks (red dashed line), (d) frequency spectrum of a residual.}
 	\label{vowel}
 \end{figure}
 
As an example, Figs. \ref{fig:one1} and Fig. \ref{fig:two1} present 100 ms of a real recording of vowel $/a/$ in the time and the frequency domain. It can be noticed that this spectrum has an harmonic structure and could be accurately parameterized by its energy, spectral envelope, and fundamental frequency. The dashed line in Fig. \ref{fig:two1}, which coincides with the spectral envelope of a vowel, represents the frequency response of the estimated LPC filter. On the other hand, the peaks of the residual signal in Fig. \ref{fig:three1} correspond to a buzzing excitation from the glottis. Finally, the frequency spectrum of a residual in Fig. \ref{fig:four1} is relatively flat and has less different formants (acoustic harmonic resonances), compared to the initial spectrum in Fig. \ref{fig:one1}. Thus, we can reach the intuitive conclusion that LPC analysis separates the spectral envelope from the harmonic content of the signal.

\begin{figure}[h]
\includegraphics[width = \textwidth]{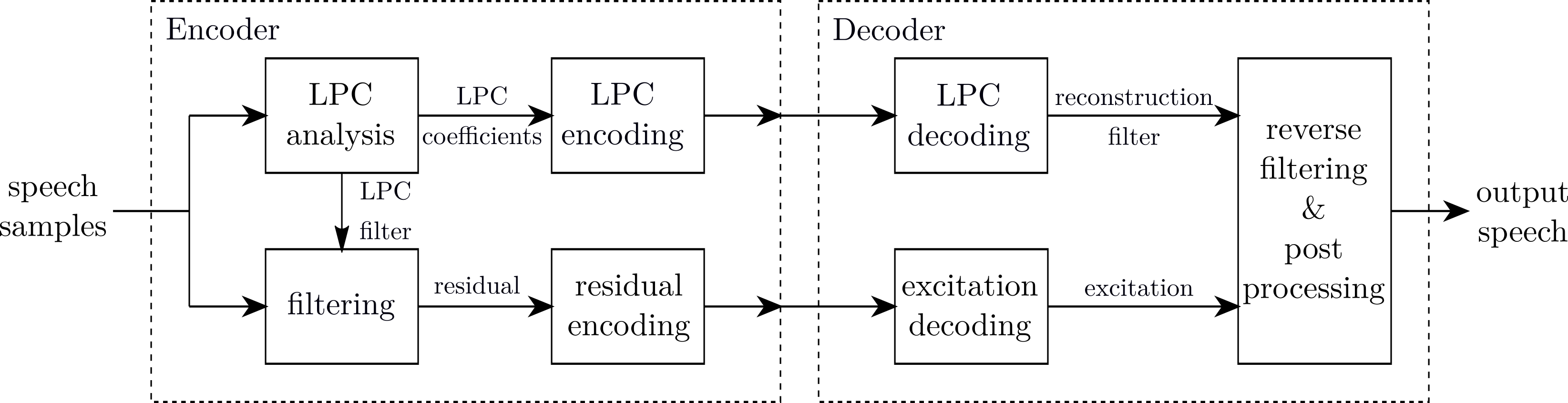}
\caption{Simplified diagram of LPC encoder and decoder.}
\label{diagram}
\end{figure}

Source-filter separation emphasizes the relevant vocal information, which is advantageous in signal compression. Figure~\ref{diagram} depicts a simplified diagram of speech analysis and synthesis by a generic LPC coder. The encoder estimates LPC coefficients and calculates the excitation of a small portion of speech (typically 5ms \textendash 20ms). Lossy excitation encoding puts stress on preserving the harmonic content of the speech, whereas LPC filters are often weighted to boost formants, taking advantage of the human auditory system's specificities and information redundancy. From this point, it is understandable that vocal parameters in a waveform are usually well preserved during compression, while the less speech-like are removed. The output waveform is also smoothed in the time and spectral domains to remove ringing effects caused by frame-based processing.

Linear Predictive Coding achieves remarkable results in representing and compressing smoothly varying voiced sounds but often struggles with encoding short and noisy plosives (like $/p/$ or $/t/$), which do not fit into the source-filter speech model. To improve the robustness for noisy sounds, LPC coders incorporate more flexibility into the excitation encoder. This observation suggests that the potential performance of the DoV technique would mostly depend on the accuracy and reaction time of excitation encoding. 

Despite preserving core speech intelligibility, time-domain LPC coding destroys the fine time-structure of compressed signals. Thus, it is not obvious how voice channels equipped with LPC coders modify the sent signal. In Section \ref{section3}, we describe a simplified framework that will allow us to evaluate the typical distortion introduced by LPC voice channels. 

\section{Data over LPC Voice Coders}
\label{section3}

This section presents a novel DoV technique based on codebooks of phase-modulated harmonic waveforms. The proposed solution is the result of extensive simulation experiments with three representative LPC narrow-band coders: AMR \citep{amr_3gpp}, Speex v1.2 \citep{speex_ietf} and Opus-Silk v1.3.1 \citep{opus_ietf}. 

The section begins with a thorough analysis of signal distortion characteristics caused by selected voice compression algorithms. The investigation leads to a significant improvement in harmonic signal demodulation. Finally, the section proposes a simplified codebook design approach.  

\subsection{Multiharmonic modulation over LPC voice coders }
\label{section3_1}

By their construction optimized to vowel sounds, LPC coders are suitable for synthesizing multi-harmonic signals. On the other hand, the versatility of excitation encoding allows easy manipulation of phase information, which above 2 kHz typically plays a lesser role in speech intelligibility \citep{rabiner2011theory,alves2014perception}. Combining phase modulation with multiple subcarriers is particularly interesting, as it opens the possibility of applying spectrally-efficient orthogonal frequency-division multiplexing (OFDM) modulation \citep{nee2000ofdm}. The OFDM approach has been already analyzed in the context of DoV in \citep{chen2011ofdm}. Their solution is based on 27 independently modulated carriers and achieved a high bitrate of 2.4 kbps over the (now obsolete) RPE-LTP GSM voice coder at an acceptably low error rate. 

\begin{figure}[h]
\centering
  \includegraphics{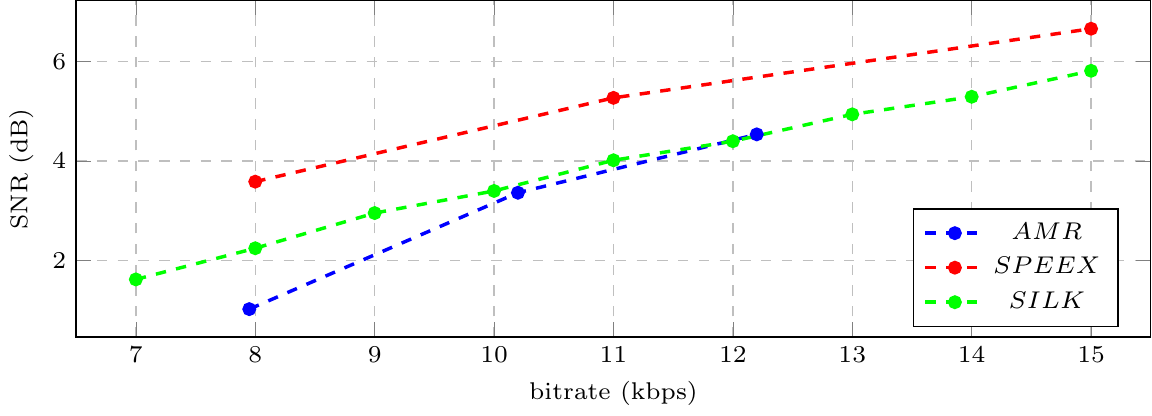}
  \caption{SNR of a multi-tone signal compressed by a selection of LPC coders. The multi-tone signal consisted of eight harmonics at frequencies 400 Hz, 800 Hz, ..., 3200 Hz with a 400~Hz step. Harmonics were independently phase-modulated with a modulation order 4 and a modulation rate of 200 baud.}
  \label{bitrate_SNR}
\end{figure}

Figure \ref{bitrate_SNR} presents a signal-to-noise ratio (SNR) of a multi-tone signal compressed by AMR, Speex, and Opus-Silk at different compression rates. It may be noticed that distortions introduced by coders are roughly similar. However, the large amount of distortion poses a big challenge for reliable data transmission, especially at compression bitrates below 10 kbps. Thus, a better understanding of the characteristics of signal distortion would help designing a more robust communication scheme. For the sake of consistency, the following experiments were performed only for fixed compression bitrates: AMR 12.2 kbs, Speex 11 kbps, and Opus-Silk 12 kbps.

\begin{figure}[!h]
\centering
  \subfloat[Radial variance of distortion.]{
{\includegraphics[width=.45\textwidth]{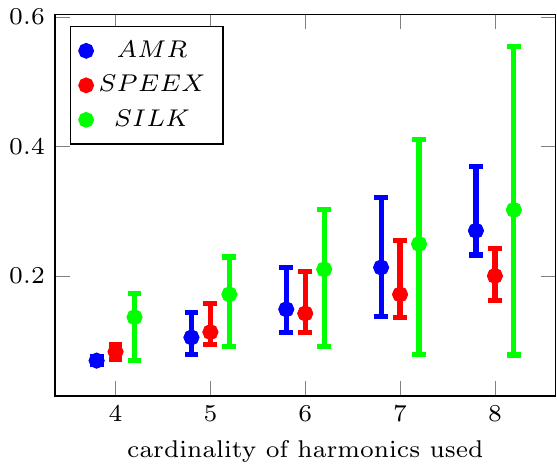}}
  }
  \hfill
 \subfloat[Error rate of phase detection.]{
{\includegraphics[width=.45\textwidth]{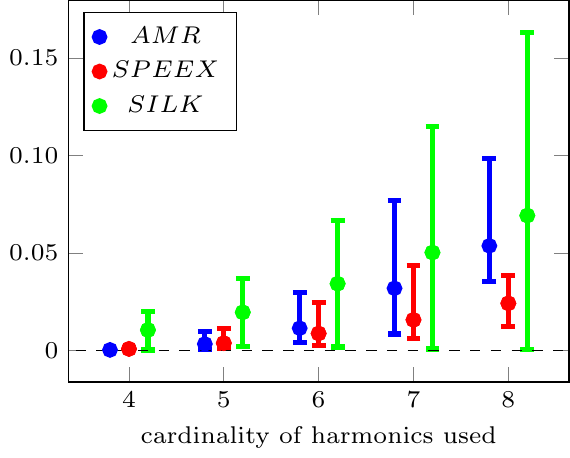}}
 }
  \caption{Energy-normalized variance of spectral distortion and related error rates of phase detection in multi-tone signals compressed by a selection of LPC coders. The initial multi-tone signal consisted of four independently phase-modulated harmonics at frequencies 400 Hz, 800~Hz, 1200 Hz, and 1600 Hz, with a modulation order 4 and a modulation rate of 200 baud. Then, the set of carriers was expanded by adding harmonics at 2000 Hz, 2400 Hz, ..., 3200 Hz with a 400~Hz step. Colored bars denote the lowest and highest values among harmonics, and dots indicate the average.}
  \label{variance}
\end{figure}

Since LPC coders process the signal jointly, it is not clear how the presence of other harmonics affects the distortion of each component. Figure \ref{variance} presents the energy-normalized variance of spectral distortion and related error rates of phase detection in multi-tone signals compressed by a selection of LPC coders. It can be noticed, that there is a direct relation between the variance of distortion and the error rate. In addition, as the cardinality of harmonics in the multi-tone signal goes up, the variability of error rates rises. Nevertheless, harmonics are not distorted uniformly, which is especially noticeable for Silk. It is because the codec puts a more significant emphasis on preserving lower frequencies \citep{opus_ietf}, especially important for the auditory perception of voice~\citep{gold2011speech}.

\begin{figure}[!h]
\centering
  \subfloat[400 Hz.]{\includegraphics[width=.22\textwidth]{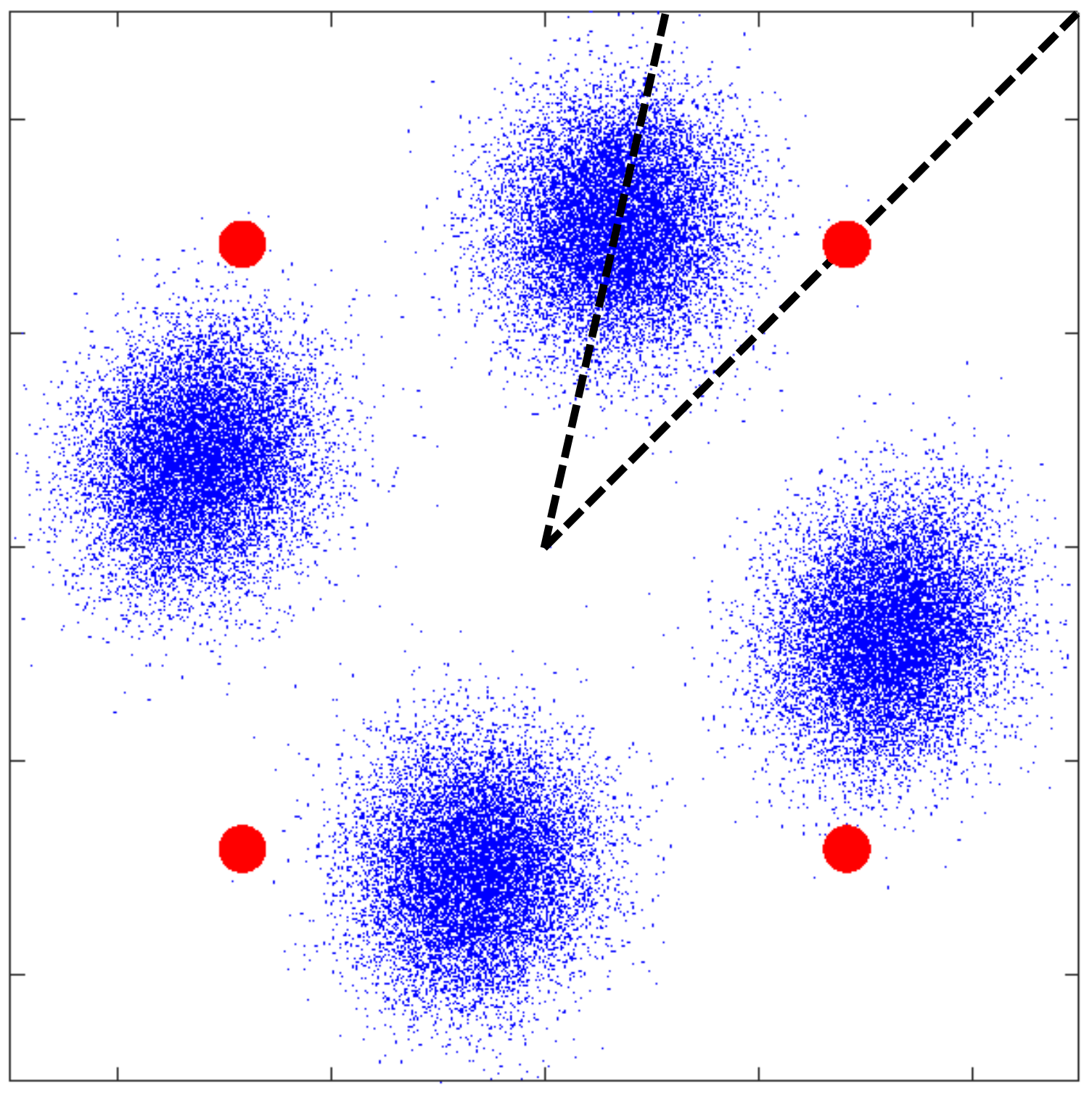}}
  \hfill
 \subfloat[800 Hz.]{\includegraphics[width=.22\textwidth]{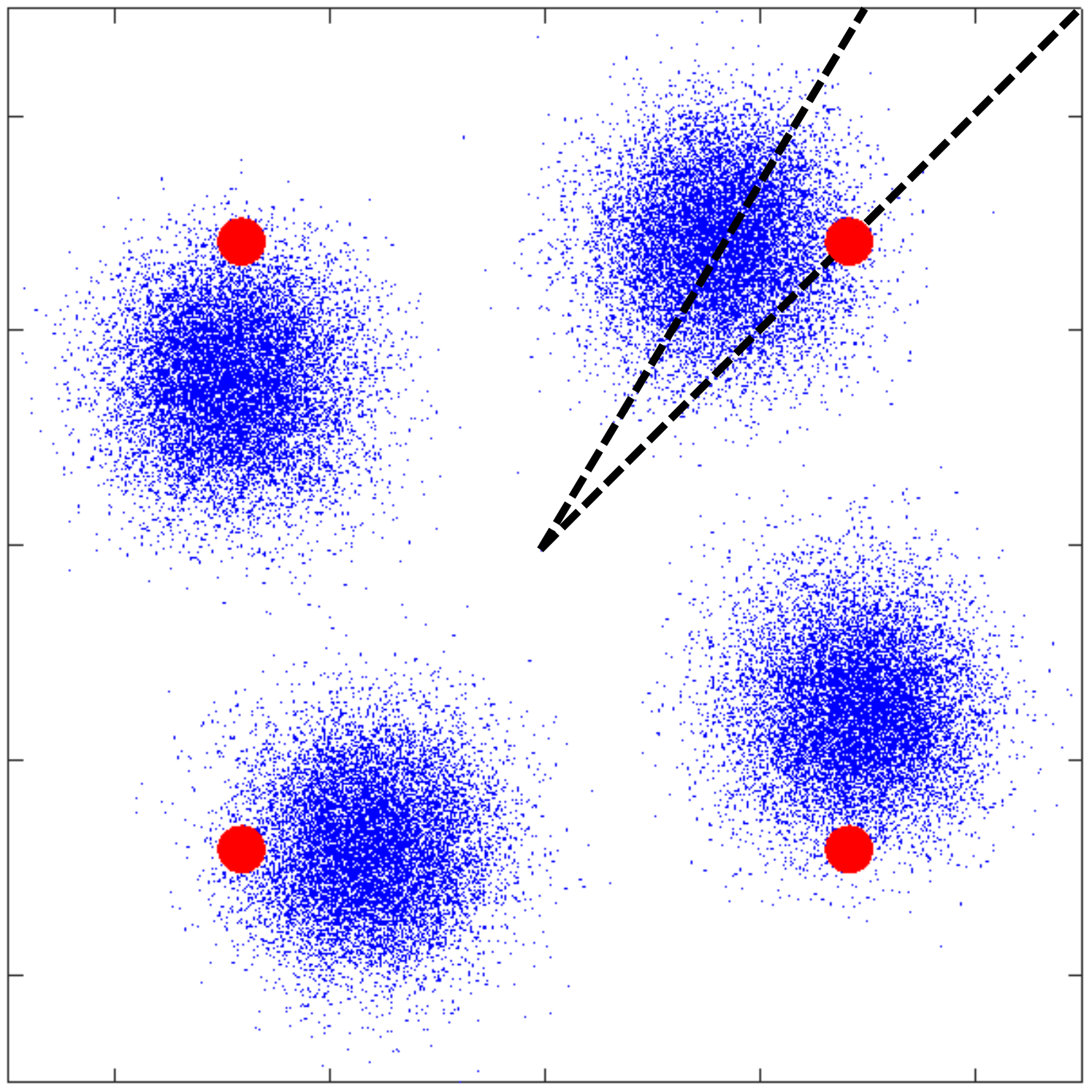}}
 \hfill
   \subfloat[1200 Hz.]{\includegraphics[width=.22\textwidth]{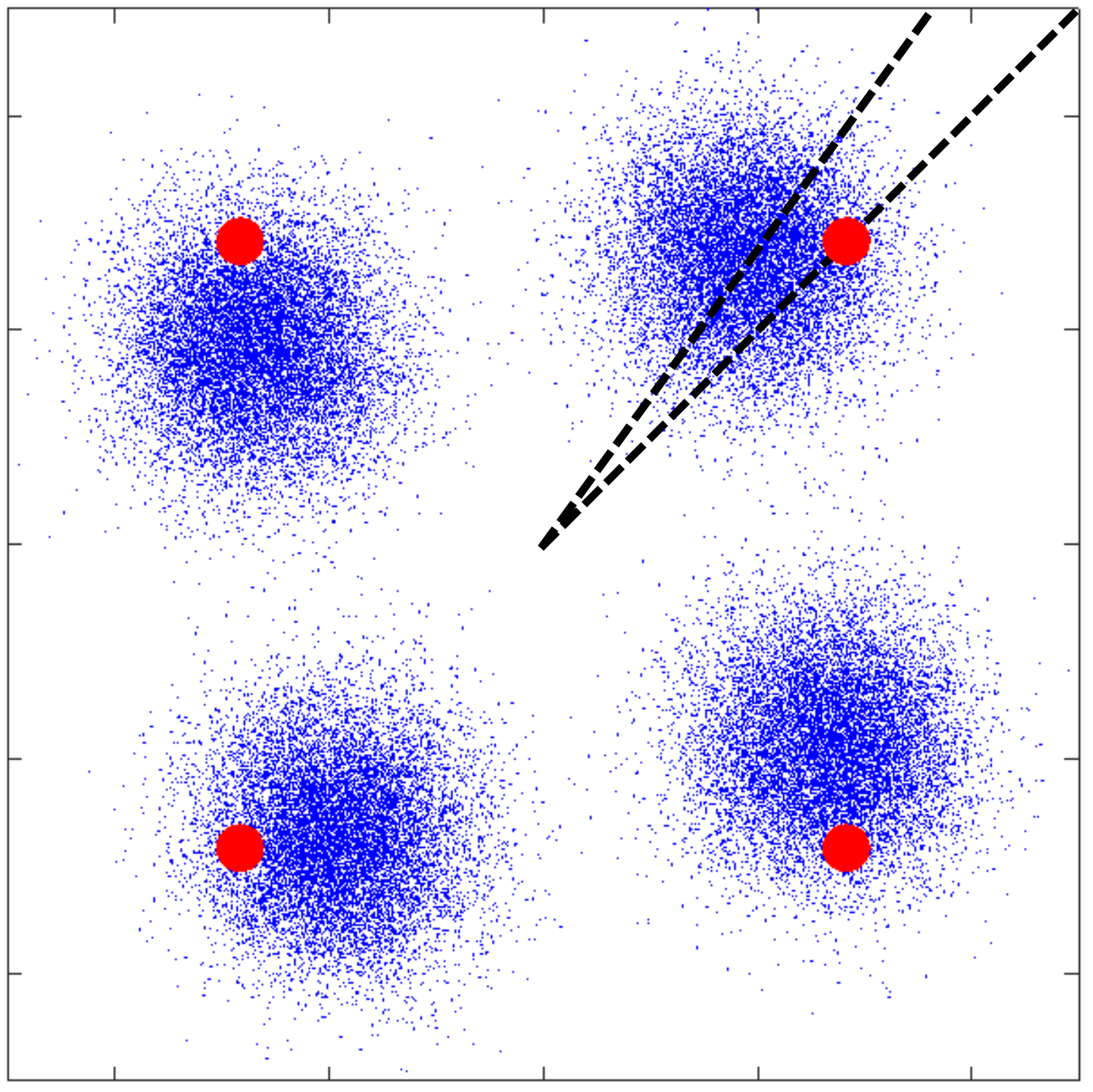}}
  \hfill
 \subfloat[1600 Hz.]{\includegraphics[width=.22\textwidth]{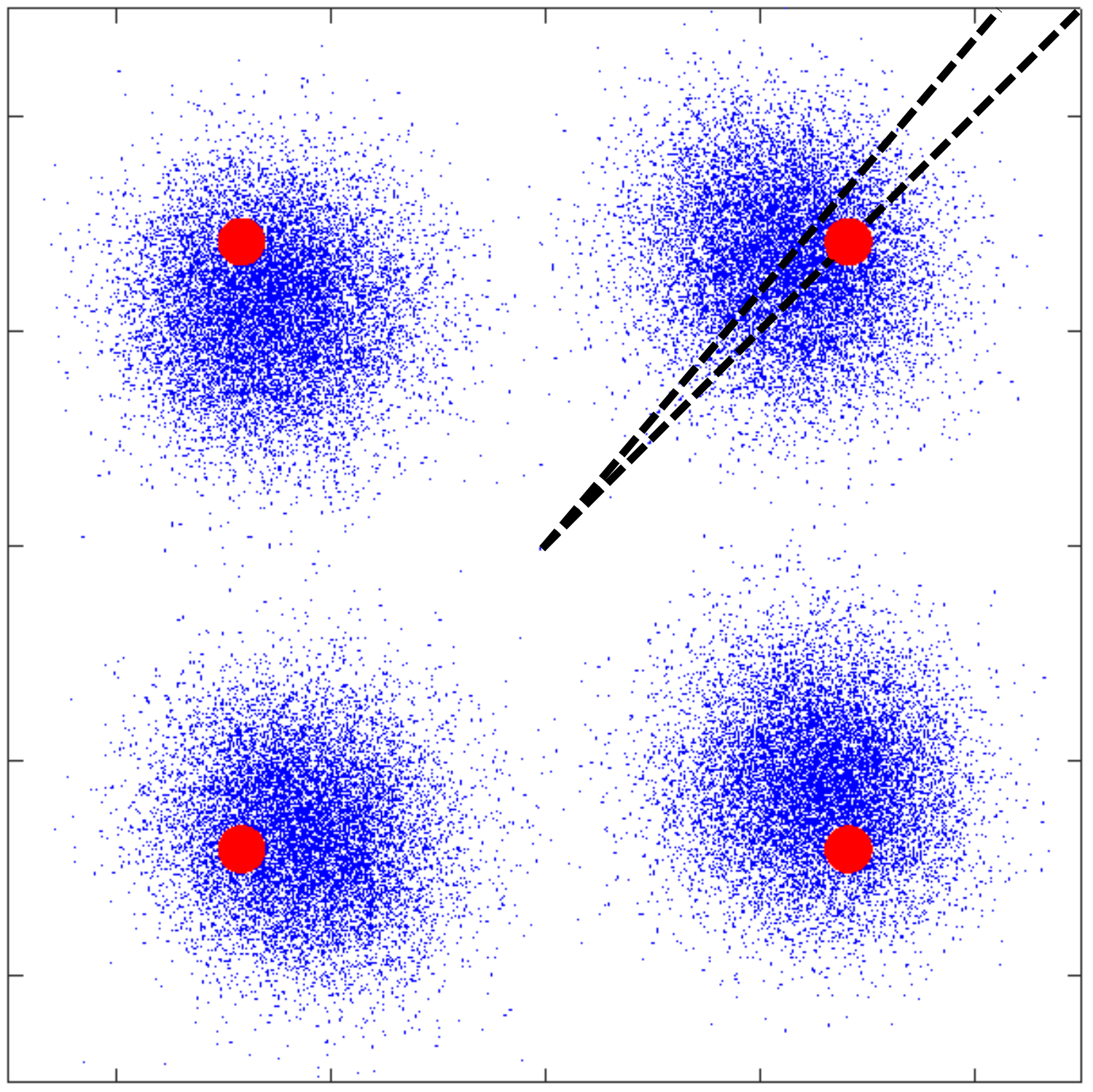}}
  \caption{Scatter plots of a four-harmonic signal compressed by AMR. Each plot represents a distortion of one phase-modulated harmonic at 400 Hz, 800~Hz, 1200 Hz, and 1600 Hz, with a modulation rate of 200 baud. Blue points correspond to compressed symbols, whereas red dots denote the initial phase constellation. The angle of the phase shift (restricted by black rays) varies in frequency.}
  \label{fdm}
\end{figure}

The distortion introduced by each studied coder has a similar nature, as presented in Fig.~\ref{fdm}. Apart from random noise-like distortion, all samples are subject to constant phase shift (this effect was also observed in \citep{lee2017vulnerability,xu2017data}). The phase shift depends on the frequency and the specific LPC coder, but not on symbol duration. The phase shift is probably introduced during speech synthesis by the LPC reconstruction filter with a non-uniform phase response.

\begin{figure}[!h]
\centering
\centering
  \subfloat[Skewness.]{
{\includegraphics[width=.42\textwidth,height=0.25\textwidth]{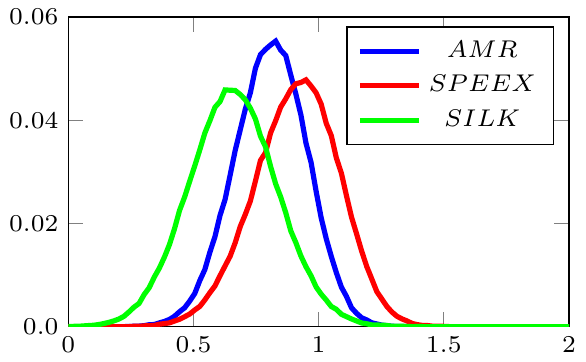}}
  }
  \hfill
 \subfloat[Kurtosis.]{
{\includegraphics[width=.42\textwidth,height=0.25\textwidth]{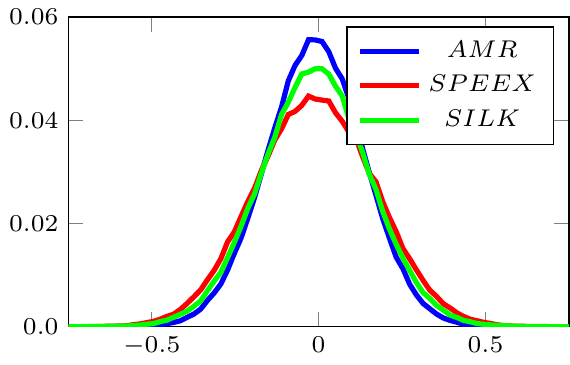}}
 }
 \\
  \subfloat[Inter-harmonic correlation.]{
{\includegraphics[width=.42\textwidth,height=0.25\textwidth]{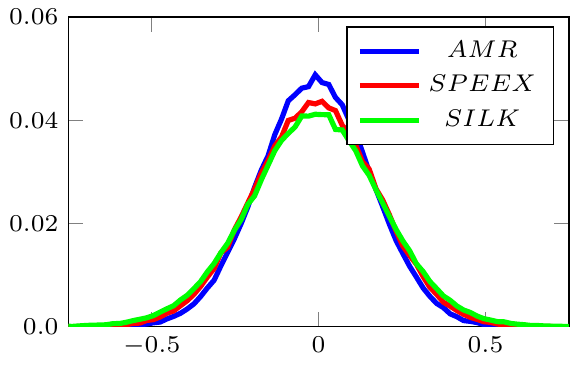}}
  }
  \hfill
 \subfloat[Time correlation.]{
{\includegraphics[width=.42\textwidth,height=0.25\textwidth]{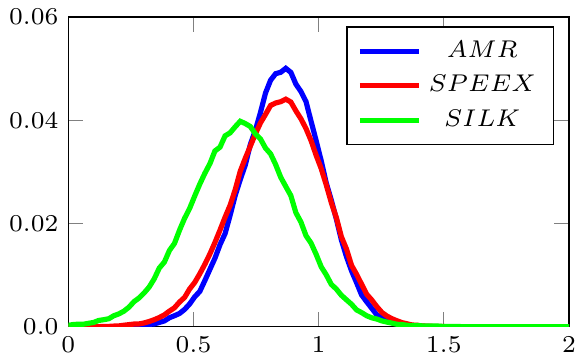}}
 }
  \caption{Sample probability density function of a variable part of distortion of two harmonics at frequencies 800~Hz and 1600 Hz, compressed by a selection of LPC coders. The compressed signal consisted of four independently phase-modulated carriers at frequencies 400 Hz, 800 Hz, 1200 Hz, and 1600 Hz, with a modulation order 4 and a modulation rate of 200 baud. Distortion in the transversal axis is centered using mean phase shift compensation, and the x-axes are normalized to the initial amplitude value of each harmonic.}
  \label{density}
\end{figure}

The sample density distributions of the variable part of distortion are approximately Gaussian, like those presented in Fig. \ref{density}. As the frequency goes up, the width (i.e., variance) is getting larger. This observation supports the intuition that the harmonics at lower frequencies are generally less distorted by compression. 

\begin{figure}[!h]
\centering
  \subfloat[Skewness.]{\label{skewness}
{\includegraphics[width=.46\textwidth]{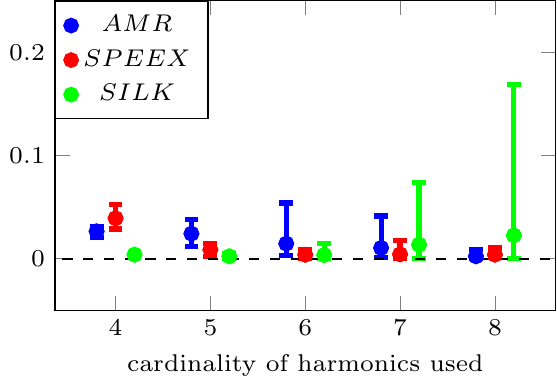}}
  }
  \hfill
 \subfloat[Kurtosis.]{\label{kurtosis}
{\includegraphics[width=.46\textwidth]{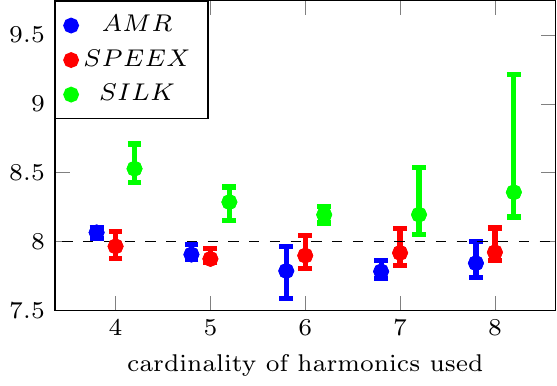}}
 }
 \\
  \subfloat[Inter-harmonic correlation.]{\label{correlation1}
{\includegraphics[width=.46\textwidth]{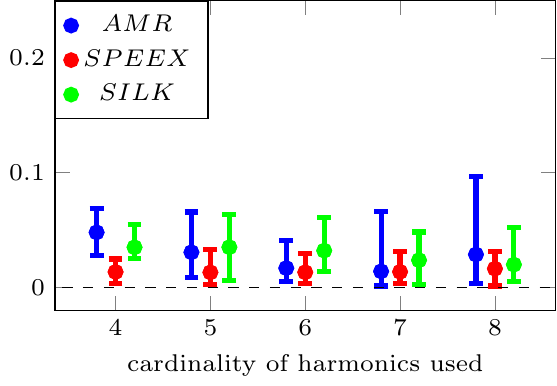}}
  }
  \hfill
 \subfloat[Time correlation.]{\label{correlation2}
{\includegraphics[width=.46\textwidth]{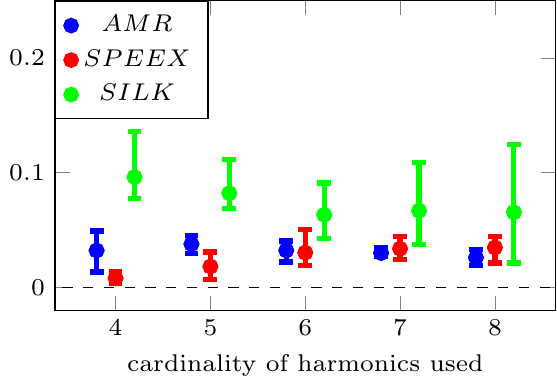}}
 }
  \caption{Statistical parameters of spectral distortion in multi-tone signals compressed by a selection of LPC coders. The initial multi-tone signal consisted of four independently phase-modulated harmonics at frequencies 400 Hz, 800~Hz, 1200 Hz, and 1600 Hz, with a modulation order 4 and a modulation rate of 200 baud. Then, the set of carriers was expanded by adding harmonics at 2000 Hz, 2400 Hz, ..., 3200 Hz with a 400 Hz step. Colored bars denote the lowest and highest values among harmonics, and dots indicate the average.}
  \label{statistics}
\end{figure}

Figures \ref{skewness} and \ref{kurtosis} present Mardia's bivariate skewness and kurtosis of a variable part of distortion. Mardia's skewness and kurtosis of a $p$-variate random sample $\mathrm{x}_1, \ ..., \ \mathrm{x}_n$ whose sample mean vector $\bar{\mathbf{x}}$ and sample covariance $S$ are defined as \citep{mardia1970measures}:

\begin{align}
\mathrm{skewness} &= \frac{1}{n^2} \sum_{k=1}^{n}\sum_{\ell=1}^{n}\left[(\mathbf{x}_k-\bar{\mathbf{x}}) \ S^{-1} \ (\mathbf{x}_\ell-\bar{\mathbf{x}})\right]^3, \\
\mathrm{kurtosis} &= \frac{1}{n} \sum_{k=1}^{n}\left[(\mathbf{x}_k-\bar{\mathbf{x}}) \ S^{-1} \ (\mathbf{x}_k-\bar{\mathbf{x}})\right]^2 \ .
\end{align}

\noindent For a sample taken from a $p$-variate normal distribution, the statistics simplify to:

\begin{equation}
\mathrm{skewness} = 0 \qquad \mathrm{and} \qquad \mathrm{kurtosis} = p \ (p+2) \ .
\end{equation}

\noindent It can be noticed that in the case of AMR and Speex (and to some extent Silk), the computed Mardia's skewness and kurtosis are close respectively to 0 and 8, which are the values characterizing symmetric bivariate normal distribution~\citep{mardia1974applications}. Crucially, distortion is not significantly correlated both in time and between harmonics (Fig. \ref{correlation1} and Fig. \ref{correlation2}). As a result, there is some evidence to treat the variable part of signal distortion as independent and memoryless. It can be seen as an advantage for demodulation but is also quite surprising because the analyzed coders are deterministic and non-linear. It suggests that distortion characteristics depend not only on LPC coders but also on statistical properties of the modulated signal.

An open question remains, though, for other LPC coders at similar compression rates. Precisely, LPC coding's basic principles do not imply the independence of distortion in time or frequency. On the other hand, it is arguable that such properties of the proposed modulation, like harmonicity and constant spectral amplitude, are compatible with LPC coding's fundamental properties. Therefore, it should be suitable for the vast majority of LPC coders. 

\subsection{DoV signal generation and demodulation}
\label{Section3_2}

\begin{figure}[h]
\centering
    \includegraphics[width=0.8\textwidth]{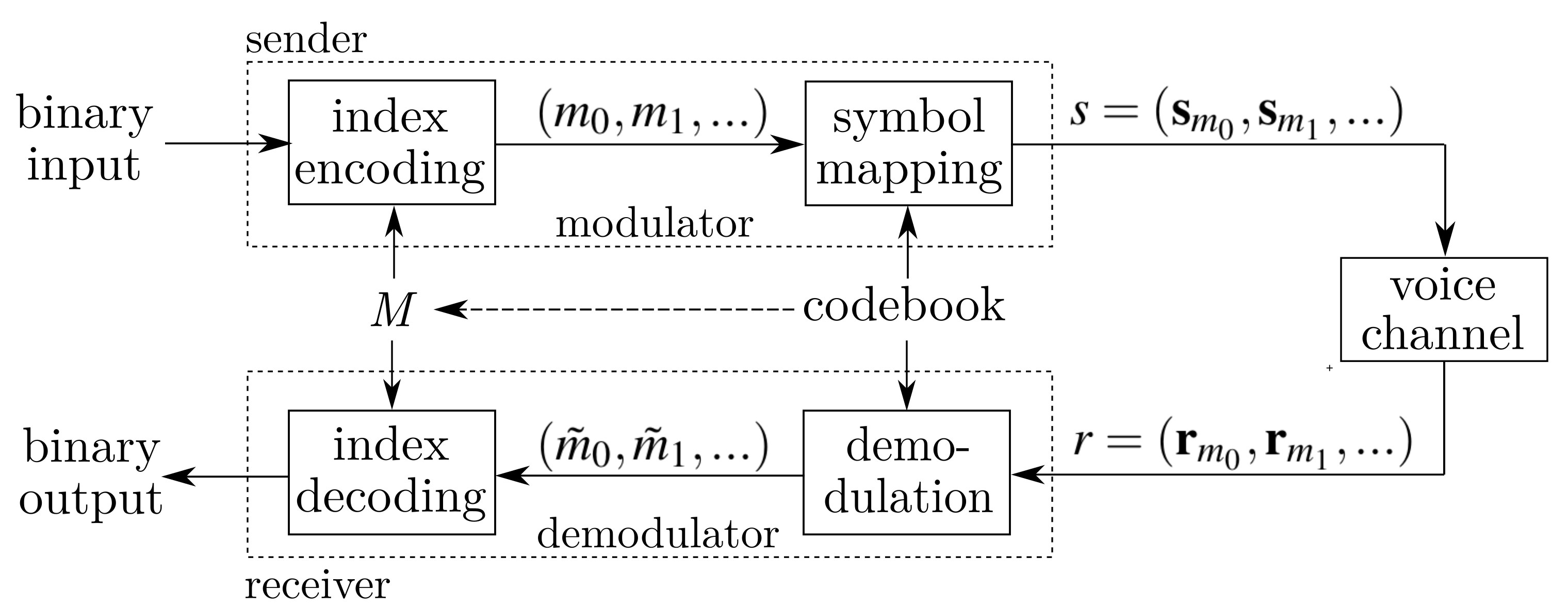}
    \caption{Modulation and demodulation of a discrete DoV signal using a codebook of $M$ pre-defined discrete audio waveforms. }
    \label{modulation}
\end{figure}

Figure \ref{modulation} depicts the typical diagram of a data transmission system over voice channel, which uses a codebook of $M$ pre-defined discrete-time audio waveforms. Signal generation is a two-step procedure that firstly encodes the binary input into a sequence of indices $(m_0, m_1,...)$ and then maps these indices into a concatenation of codebook symbols $s = (\mathbf{s}_{m_0}, \mathbf{s}_{m_1},...)$. Finally, the resulting discrete-time audio signal $s$ is played to the (digital) audio input of a voice channel. 

On the reception side, the demodulator splits the received sampled audio signal $r = (\mathbf{r}_{m_0}, \mathbf{r}_{m_1},...)$ into short chunks of fixed length corresponding to the symbol duration, and then performs symbol-by-symbol matched-filtering with all codebook entries. In the last steps, the demodulator extracts the indices of the codebook symbols giving the highest correlation value and decodes the binary information.

\begin{figure}[h]
\centering
 \subfloat{
{\includegraphics[width=1\textwidth]{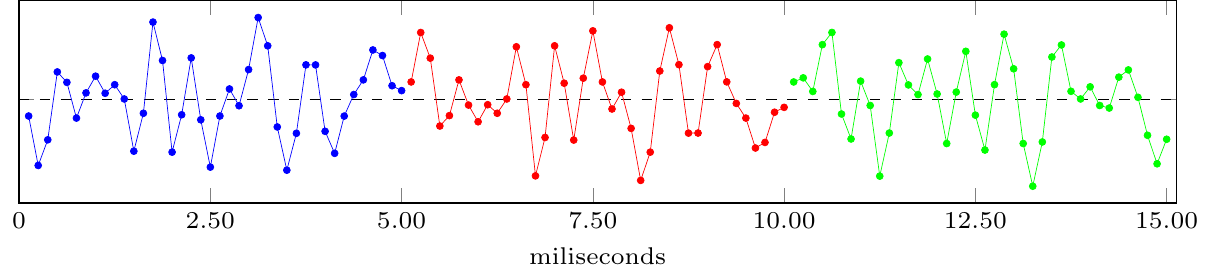}}
 }
  \caption{Three discrete-time codebook waveforms (respectively blue, red and green dots) of duration 5 ms and consisting of 10 harmonics at frequencies 600 Hz, 800 Hz, ..., 2400 Hz, with a 200 Hz step.}
  \label{DoV_symbol}
\end{figure}

In the proposed DoV technique, a codebook symbol is a vector of waveform samples $\mathbf{s}_m = [s_m[0],..., s_m[N-1]]$ sampled at 8 kHz and of duration between 2.5-10 ms. Each symbol consists of some small number $K$ (between 7-10) of orthogonal harmonics modulated by quadrature phase-shift keying (4-PSK): 

\begin{equation}
s_m[n] = \mathrm{Real} \left( \sum_{k=0}^{K-1}C_{m,k}\exp \left(j(k+k_0)\omega_0\frac{n}{N}\right) \right), \quad n = 0, 1, ..., N-1, 
\end{equation}

\noindent where $0\leq m< M$ is the symbol index, $\omega_0 $ denotes the fundamental angular frequency and $k_0$ is the subband of the lowest harmonic. Finally, $\mathbf{C}_m = \{C_{m,k} \ | \ 0 \leq k < K\}$ denotes a sequence of $K$ complex PSK symbols over the phase-amplitude plane: 

\begin{equation}
C_{m,k} = A \cdot \exp(j2\pi\varphi_{m,k}/4), \quad k = 0,...,K-1,
\end{equation}

\noindent where $A$ is the amplitude and $\Phi_m = \{2\pi\varphi_{m,k}/4 \ | \ 0 \leq k < K, \ \varphi_{m,k} \in \mathbb{Z}_4\}$ denotes a sequence of PSK phases (the selection of phase sequences will be detailed in Section~\ref{section3_3}). Examples of such waveforms are presented in Fig. \ref{DoV_symbol}.  

The symbol structure is equivalent to the discrete-time base-band representation of 4PSK-OFDM modulation \citep{nee2000ofdm}. Therefore, the received symbols can be processed in a similar manner using subband de-multiplexing. Let $\mathbf{\tilde{C}}_m = \{\tilde{C}_{m,k} \ | \ 0 \leq k < K\}$ be the sequence of PSK symbols obtained from some received codebook symbol $\mathbf{r}_m$. Assuming a typical AWGN (Additive White Gaussian Noise) channel, the maximum likelihood OFDM symbol detection can be expressed by the $\mathrm{L}^2$ norm minimization in the complex plane \citep{schulze2005theory}:

\begin{equation}
\tilde{m} = \arg \min_m \ \sum_{k=0}^{K-1} \left| \tilde{C}_{m,k} -  A\exp\left(j2\pi\varphi_{m,k}/4\right) \right|^2.
\label{min_eucl}
\end{equation}

\noindent However, the experiments in Section \ref{section3_1} indicate that compression by the selected LPC coders causes group delay in the processed signal and alters each harmonic with a distortion of different variance. The estimated phase shift $\hat{\phi}_k$ and the variance of distortion $\hat{\sigma}^2_k$ respective to each harmonic can be computed using a training sequence and the following estimators for sample mean and sample variance \citep{statisticsbook}:
\begin{align}
\hat{\mu}_k &= |\hat{\mu}_k|\exp\left( j\hat{\phi}_k \right) = \frac{1}{L}\sum_{\ell=0}^{L-1} \tilde{C}_{m_\ell,k}\exp(-j2\pi\varphi_{m_\ell,k}/4) ,\\
\hat{\sigma}^2_k &= \frac{1}{L-1}\sum_{\ell=0}^{L-1} \left| \tilde{C}_{m_\ell,k}\exp(-j2\pi\varphi_{m_\ell,k}/4) - \hat{\mu}_k \right|^2,
\end{align}

\noindent where $\mathbf{\tilde{C}}_{m_\ell} = \{\tilde{C}_{m_\ell,k} \ | \ 0 \leq k < K\}$ denotes the $\ell-th$  sequence of PSK symbols measured at the reception side and $\Phi_{m_\ell} = \{2\pi\varphi_{m_\ell,k}/4 \ | \ 0 \leq k < K, \ \varphi_{m_\ell,k} \in \mathbb{Z}_4\}$ denotes the initial phases of the corresponding codebook symbols in the training sequence.  

With the estimated $\hat{\phi}_k$ and $\hat{\sigma}^2_k$, one may apply the phase shift compensation and spectral weighting of distortion in the demodulation rule from Eq. (\ref{min_eucl}):
\begin{equation}
\tilde{m} = \arg \min_m \ \sum_{k=0}^{K-1} \left| \mathbf{\tilde{C}}_k\exp(-j\hat{\phi}_{k}) -  A\exp\left(j2\pi\varphi_{m,k}/4\right) \right|^2/\hat{\sigma}_k^2.
\label{min_eucl_cor}
\end{equation}

\noindent Finally, rewriting Eq. (\ref{min_eucl_cor}) and removing the constant terms gives a more convenient demodulation rule, which is maximizing the real part of a complex dot product \citep{schulze2005theory}: 
\begin{equation} 
\tilde{m} = \arg \max_m \ \mathrm{Real} \left(\sum_{k=0}^{K-1}\mathbf{\tilde{C}}_k \cdot  \ \frac{A}{\hat{\sigma}_{k}^2}\exp\left(-j2\pi\varphi_{m,k}/4-j\hat{\phi}_{k}\right)\right).
\label{demequationcorr}
\end{equation}

In contrast to time-domain matched-filtering, the proposed demodulation rule enables phase and variance correction in the channel distortion. Secondly, it becomes more efficient when the codebook size grows. Instead of performing $M$ matched-filtering operations on a symbol of length $N$, this demodulator needs to compute the in-phase/quadrature (I/Q) representations of $K < N$ PSK symbols and to correlate them with $M$ different phase sequences. As an example, given the triple $(K,M,N) = (8,256,40)$, matched filtering in the time domain requires at least $256\cdot 40 = 10 \ 240$ real-value multiplications. On the other hand, demodulation using Eq. (\ref{demequationcorr}) involves computing the complex PSK symbols ($2 \cdot 8\cdot 40 = 640$ real-value multiplications) and comparing the obtained sequence with all phase combinations in the codebook ($8 \cdot 256 = 2048$ complex multiplications, or at least $4096$ real-value multiplications). 

Despite the computational improvement, the codebook's preferable size ranges between 64 and 256 elements and should not overreach $4096$ elements. These values would make the real-time demodulation computationally practical on portable devices, especially if the codebook has a symmetric structure that enables further computational optimizations. 

Another factor in the process of selecting the codebook size is the transmission bitrate. Full 4PSK-OFDM modulation offers transmission up to $2K = \log_2(4^K)$ information bits per symbol. However, the modulation is susceptible to excessive distortion or attenuation of some harmonics in spectrally selective voice channels. Instead, it is advisable to choose only a subset of all possible OFDM phase combinations to enlarge the minimum distance between symbols. This approach makes a transmission over voice channels more robust to spectrally selective distortion, as a large distortion of some harmonics would be compensated by a moderate distortion of the others. On the other hand, smaller modulation order $M < 4^K$ decreases the bitrate. 

\subsection{Codebook design}
\label{section3_3}

Construction of a suitable DoV codebook relies on finding (or training) a subset of harmonic symbols with a large minimum distance. However, this task becomes challenging as the number of symbol combinations increases. This subsection gives a proposition of a suboptimal codebook design method, which produces a set of harmonic waveforms sufficiently different from each other.

For $\mathbf{x}_1,\mathbf{x}_2 \in \mathbb{C}^K$, let $\mathrm{d}_{E}(\mathbf{x}_1,\mathbf{x}_2)$ be the Euclidean metric over the complex space and for $\mathbf{y}_1,\mathbf{y}_2 \in \mathbb{Z}_4^K$, let $\mathrm{d}_L(\mathbf{y}_1,\mathbf{y}_2)$ be the Lee metric over $\mathbb{Z}_4^K$:
$$\mathrm{d}_{L}(\mathbf{y}_1,\mathbf{y}_2) = \sum_{k=0}^{K-1}\min(|y_{1,k}-y_{2,k}|, \ 4-|y_{1,k}-y_{2,k}|). $$ 
In addition, let us define the bijective function $f: \ \mathbb{C}^K \rightarrow \mathbb{Z}_4^K $ which takes the phase indices $\varphi_{m,k}$ of every \mbox{4-PSK} sequence $\mathbf{C}_m = \{A\cdot \exp(j2\pi\varphi_{m,k}/4) \ | \ 0 \leq k < K, \ \varphi_{m,k} \in \mathbb{Z}_4 \},$ and maps to a quaternary codeword $f(\mathbf{C}_m) = \{\varphi_{m,k} \ | \ 0 \leq k < K\}$ over~$\mathbb{Z}_4^K$. For any two 4-PSK sequences $\mathbf{C}_{m_1}$ and $\mathbf{C}_{m_2}$,  we get an isometric property:
$$2A^2\mathrm{d}_{L}(f(\mathbf{C}_{m_1}),f(\mathbf{C}_{m_2})) = \ \mathrm{d}_{E}^2(\mathbf{C}_{m_1},\mathbf{C}_{m_2}).$$ 

It can be noticed that the same relation holds for the minimum distance between all PSK sequences in the OFDM codebook and elements of the associated quaternary codewords. The selection of the most distinct OFDM symbols could be thus replaced by the construction of a quaternary code $\mathcal{C} \subset \mathbb{Z}_4^K$ (not necessarily a subgroup), that maximizes the minimum Lee distance. 

In the perspective of non-binary codes with a defined minimum distance, these OFDM symbols can be seen as error correcting codes encoded in the spectral domain~\citep{wilkinson1995minimisation}. In consequence, quaternary codes provide a new degree of freedom in the DoV codebook design. By some sensible manipulation of the number of harmonics $K$, the symbol duration $N$, and the minimum distance between codebook symbols $d$, it is possible to find a codebook providing the required bitrate and maintaining sufficient robustness to distortion. Moreover, the codebook generation is computationally constrained mostly by finding quaternary codes, which is a much faster process compared to training a full codebook of waveforms. Finally, quaternary codes can be reused to produce waveforms of different duration and harmonic frequencies. It is also worth noticing that the above motivation for exploiting non-binary codes is slightly different from other works focusing mainly on reducing the peak-to-mean energy ratio of the OFDM signal \citep{davis1999peak,chen2007combined,ginige2001dynamic,hisojo2014low}. 

Due to some rotational symmetries of quaternary codes, there is no unique codebook with the largest minimum distance. It gives more flexibility in the fine-tuning of the codes to make them more suitable in real operation. It is advisable to select a codebook with a possibly uniform distribution of phase values and remove symbols with the highest maximum amplitude. Table \ref{quaternary_codes} presents the minimum distance of several quaternary codes found by a greedy Algorithm \ref{procedure}. The subroutine $\texttt{ChooseInitial}$ inserts a random or some pre-defined initial codeword into the codebook, while the subroutine $\texttt{SelectCodeword}$ iteratively selects a codeword to remain within the uniform distribution of phase values in the expanded set. 

To improve the computational demodulation efficiency, one may exploit the reflection symmetry of the codebook produced by the algorithm. Since for any $0 \leq 2m < M$ we have $\mathbf{s}_{2m} = -\mathbf{s}_{2m+1}$, it is sufficient to correlate the received PSK sequence only with codebook symbols having the even indices and then to check the sign of computation.

\begin{algorithm2e}
\caption{CodebookSearch($C, \ M$)}
\SetAlgoLined
\KwData{the set of quaternary codewords $\mathrm{C}$, an even size of codebook $M$;}
\KwResult{a set $\mathrm{Cb}$ of $M$ quaternary codes;}
$\mathrm{Cb} \longleftarrow \varnothing$\;
\tcp*[h]{select the first codeword (random or pre-defined)}\

$c_0 \longleftarrow \mathrm{ChooseInitial}(\mathrm{C})$\; 
$\mathrm{Cb} \longleftarrow \mathrm{Cb} \ \cup \ \{c_0, -c_0\}$\;
\For{$i\leftarrow 1$ \KwTo $\lfloor M/2 \rfloor-1$}{
\tcp*[h]{select codewords in $\mathrm{C}$ with a maximum Lee distance from $\mathrm{\mathrm{Cb}}$}\

$\mathrm{S} \longleftarrow \mathrm{MaxLeeDistance}(\mathrm{C}, \ \mathrm{Cb})$\;
\tcp*[h]{select a codeword from $\mathrm{S}$ respective to uniform distribution}\ 

$c_{2i} \longleftarrow \mathrm{ChooseCodeword}(\mathrm{S}, \ \mathrm{Cb})$\; 
$\mathrm{Cb} \longleftarrow \mathrm{Cb} \ \cup \ \{c_{2i}, -c_{2i}\}$\;
}
\label{procedure}
\end{algorithm2e}
\begin{table}[h]
\centering
\caption{Minimum Lee distance of additive quaternary codes of length $n = 7,8,9$ and $10$, found by Algorithm \ref{procedure}. Parameter $k$ denotes the number of (quaternary) information bits of the code. From the perspective of OFDM symbols, value $n$ is related to the cardinality of harmonics, while $k$ describes the codebook size equal to $4^k$.}
\begin{tabular}{|l|c c c c c c c c c c c c c|}
\hline
$n \setminus k$ & 2.0 & 2.5 & 3.0  & 3.5  & 4.0  & 4.5  & 5.0  & 5.5  & 6.0 & 6.5 & 7.0 & 7.5 & 8.0  \\    \hline \hline
7 & 6 & 6 & 4  & 4  & 3  & 3  & 2  & 2  & 2  & 1  & 1 & - & -  \\  
8 & 8 & 8 & 6  & 6  & 4  & 4  & 4  & 4  & 2  & 2  & 2 & 2 & 1  \\    
9 & 8 & 8 & 6  & 6  & 5  & 4  & 4  & 4  & 3  & 2  & 2 & 2 & 1  \\    
10& 10 & 9 & 7  & 6  & 6  & 5  & 5  & 4  & 4  & 3  & 3 & 2 & 2  \\    \hline
\end{tabular}
\label{quaternary_codes}
\end{table}

\section{Experiments}
\label{Section_experiments}

This section presents the performance results of the DoV scheme described in Section~\ref{section3}. Simulations are followed by experimental tests over 3G and VoIP. Examples of some DoV signals recorded during tests are available online.\footnote{\url{https://github.com/PiotrKrasnowski/Data\_over\_Voice}}

\subsubsection{Channel estimation}

\begin{figure}[h!]
\centering
  \subfloat[Standard error of phase-shift estimation.]{
 {\includegraphics[width=.45\textwidth]{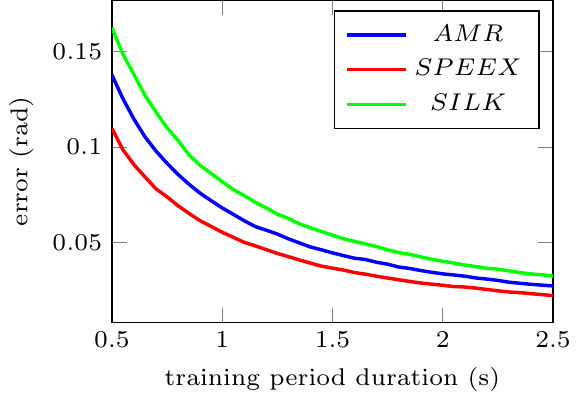}}
  }
  \hfill
 \subfloat[Standard error of variance estimation.]{
 {\includegraphics[width=.45\textwidth]{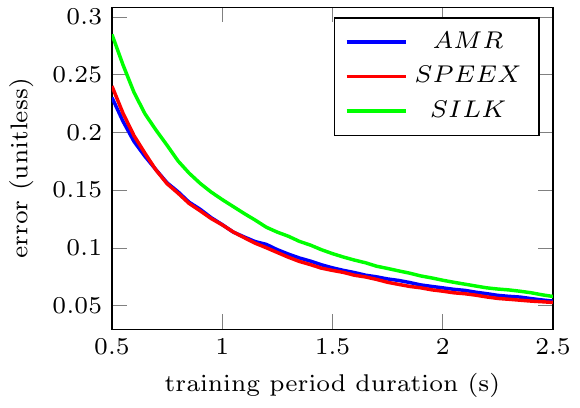}}
 }
  \caption{Estimated standard error of the phase-shift $\hat{\phi}_k$ and the normalized variance $\hat{\sigma}_k^2/\bar{\sigma}^2_{k}$ estimators of distortion introduced by a selection of coders. The graphs present the maximum standard error over all harmonics~$k$, and for every $t$ between 0.5 and 2.5 seconds with a 0.05 second step. Results obtained based on 1000 Monte Carlo runs. The reference values $\bar{\phi}_{k}$ and $\bar{\sigma}^2_{k}$ were computed from a sample of 50 000 symbols. The DoV signal consisted of 8 harmonics at frequencies 400 Hz, 800 Hz, ..., 3200 Hz with a modulation rate of 200 baud.}
  \label{sem}
\end{figure}

Efficient detection of received DoV symbols, described by Eq. (\ref{demequationcorr}) in Section \ref{section3}, requires voice channel characterization using the training sequence. Intuitively, the larger number of symbols in the sequence, the more accurate is the estimation. We estimated the standard error ($\mathrm{SE}$) of the phase shift $\hat{\phi}_k(t)$ and the variance of distortion $\hat{\sigma}^2_k(t)$ as a function of training duration $t$, using Monte Carlo simulations and the following formulas:
\begin{align}
\mathrm{\widehat{SE}}^2_{\hat{\phi}_k(t)} &= \frac{1}{L}\sum^{L}_{\ell = 1}\left(\hat{\phi}_{k,\ell}(t) - \bar{\phi}_{k} \right)^2,\\
\mathrm{\widehat{SE}}^2_{\hat{\sigma}^2_k(t)/\bar{\sigma}^2_{k}} &= \frac{1}{L}\sum^{L}_{\ell = 1}\left(\hat{\sigma}^2_{k,\ell}(t) - \bar{\sigma}^2_{k} \right)^2/\bar{\sigma}^2_{k},
\end{align}

\noindent where $L$ is the number of Monte Carlo runs, $\hat{\phi}_{k,\ell}(t)$ and $\hat{\sigma}^2_{k,\ell}(t)$ denote respectively the estimated phase shifts and the variances of distortion in the $\ell -th$ Monte Carlo run, and the reference values $\bar{\phi}_{k}$ and $\bar{\sigma}^2_{k}$ were obtained from a sequence of 50 000 DoV symbols (250 seconds of a signal). Figure~\ref{sem} depicts the maximum standard error of $\hat{\phi}_k(t)$ and $\hat{\sigma}^2_k(t)/\bar{\sigma}^2_{k}$ taken over all harmonics~$k$ and for every $t$ between 0.5 and 2.5 seconds with a 0.05 second step. It can be observed that 2 seconds of training period should give a sufficiently accurate channel characterization. 

\subsection{Simulations}

\begin{figure}[]
\centering
  \subfloat{\label{ser1}
 {\includegraphics[width=1\textwidth]{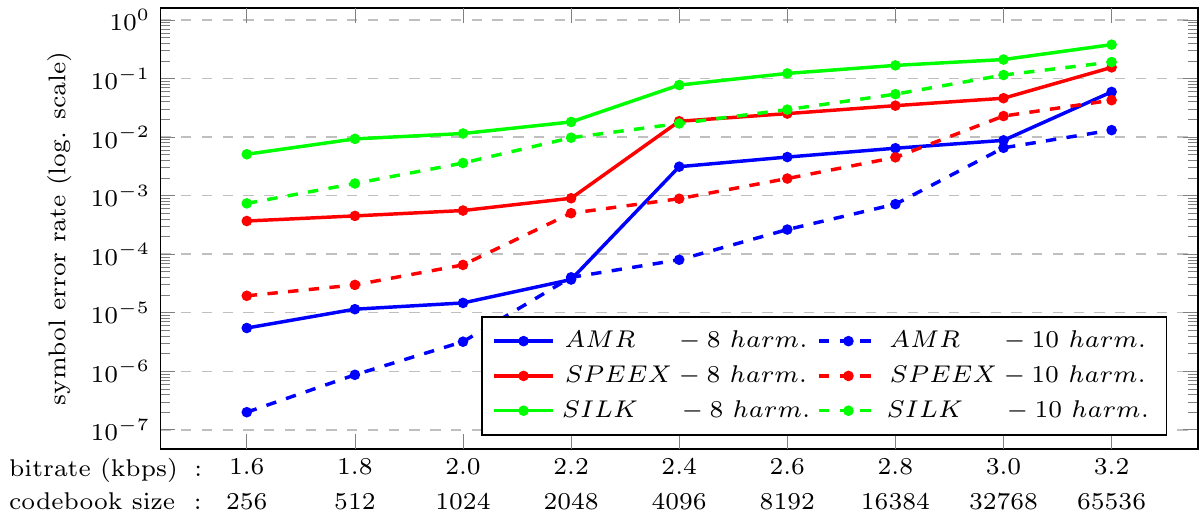}}
  }
  \\
 \subfloat{\label{ser2}
 {\includegraphics[width=1\textwidth]{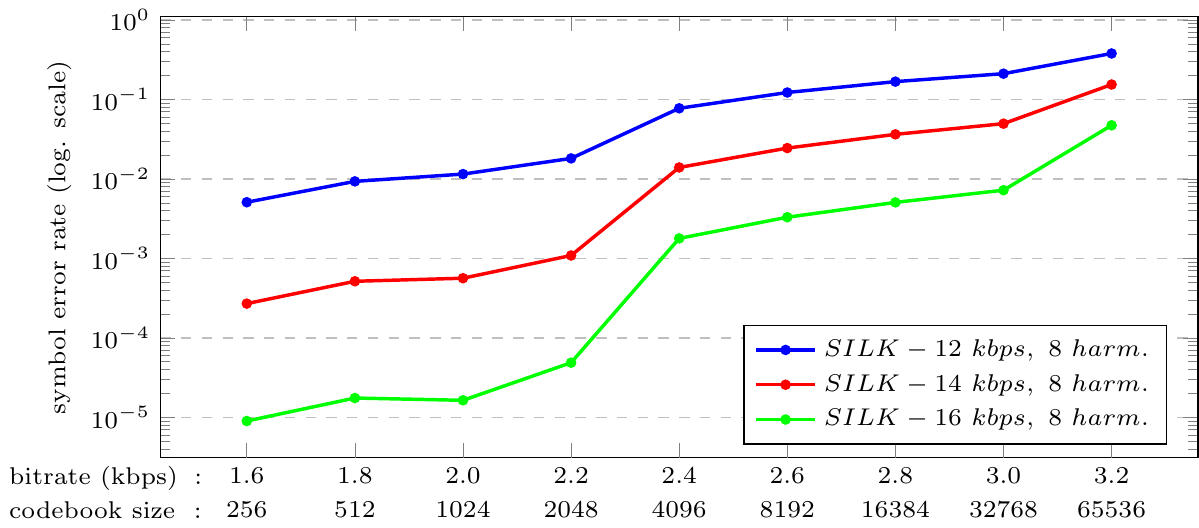}}
 }
 \\
  \subfloat{\label{ser3}
 {\includegraphics[width=1\textwidth]{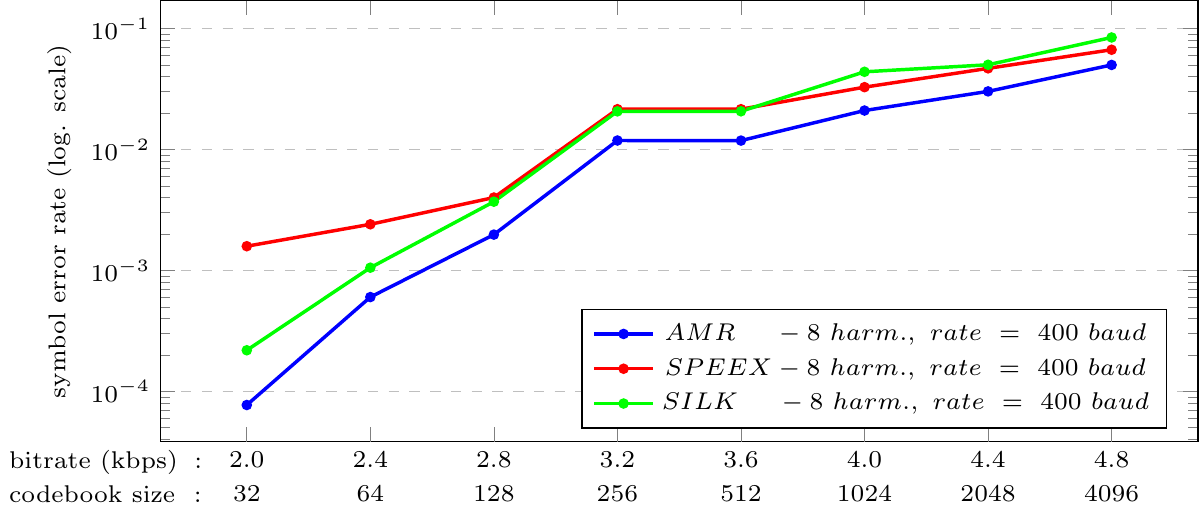}}
  }
  \caption{Decoding symbol error rate of a DoV signal compressed by AMR, Speex and Opus-Silk. To ensure reliability of the simulations, duration of the training period was extended to 4 seconds. If not indicated otherwise, symbol rate equals 200 baud. DoV signals consisted of $10^7$ symbols produced according to an output of a built-in pseudo-random generator with a pre-defined seed.}
  \label{SER}
\end{figure}

The symbol error rate primarily depends on the distortion variance and the minimum distance between codebook symbols. For example, it can be noticed in Fig.~\ref{ser1} that compressing by AMR leads to significantly lower error rates when compared to compression using the Silk codec. This result agrees with the experimental outcomes shown in Fig.~\ref{variance} in Section \ref{section3}. Nevertheless, when the voice channel's capacity goes up, the amount of distortion, and thus the error rate gradually decreases, as indicated by Fig. \ref{ser2}. 

The characteristic staircase shape of the graphs in Figs. \ref{ser1} and \ref{ser2} corresponds to the codebook minimum distance $d$ in function of the codebook size (ref. Table~\ref{quaternary_codes}). Thus, the symbol error rates obtained can be viewed as the approximated probability of the signal distortion exceeding the distance $d/2$. Consequently, it is generally advantageous to design the codebook with a larger number of orthogonal harmonics, leading to increased minimum distance and improved robustness.

Despite its simplicity, the presented scheme suffers from the large size of the codebooks used, especially at higher bitrates. The exponentially growing number of correlations becomes a major practical limitation for real-time signal demodulation. The problem can be tackled by scaling down the symbol duration at the expense of higher relative distortion and a smaller number of orthogonal frequency slots. As shown by Fig. \ref{ser3}, a modulation based on smaller codebooks of shorter symbols provides similar performance at a much lower computational cost.

\subsection{Real-world tests}
\label{real_section}

The DoV technique has been tested over a real voice channel between mobile phones, using pre-computed DoV signals. The selected phones for experiments were two iPhones 6 running iOS 12 and a Huawei P8 Lite running Android 8, each registered to a different major French mobile network operator. The DoV performance over 3G calls is displayed in Table \ref{ser_3G}, and the performance over VoIP calls using 4G wireless network is shown in Table \ref{ser_VoIP}. The duration of the training period was extended to 4 seconds to ensure the reliability of the experiments.

\begin{table}[h!]
\caption{Symbol error rate of DoV signal over 3G call with and without channel estimation.}
\parbox{.48\linewidth}{
\centering
\begin{tabular}{|c|cc|}
\hline
 \multicolumn{3}{|c|}{10 harmonics, symbol duration 5 ms} \\ \hline
bitrate & 4 s training period  & no training  \\ \hline
1.0 kbps & $< 1.0 \cdot 10^{-4}$  & $< 1.0 \cdot 10^{-4}$ \\
1.2 kbps & $< 1.0 \cdot 10^{-4}$ & $< 1.0 \cdot 10^{-4}$ \\
1.4 kbps & $1.2 \cdot 10^{-4}$ & $2.9 \cdot 10^{-4}$ \\
1.6 kbps & $2.6 \cdot 10^{-4}$ & $4.8 \cdot 10^{-4}$ \\
1.8 kbps & $6.0 \cdot 10^{-4}$ & $1.5 \cdot 10^{-3}$  \\
2.0 kbps & $1.2 \cdot 10^{-3}$ & $2.6 \cdot 10^{-3}$ \\
2.2 kbps & $9.4 \cdot 10^{-3}$ & $1.4 \cdot 10^{-2}$ \\
2.4 kbps & $1.6 \cdot 10^{-2}$ & $2.2 \cdot 10^{-2}$ \\ \hline
\end{tabular}
}
\hfill
\parbox{.48\linewidth}{
\centering
\begin{tabular}{|c|cc|}
\hline
\multicolumn{3}{|c|}{8 harmonics, symbol duration 2.5 ms} \\ \hline
bitrate & 4 s training period  & no training \\ \hline
1.2 kbps & $< 1.0 \cdot 10^{-3}$ & $< 1.0 \cdot 10^{-3}$ \\
1.6 kbps & $< 1.0 \cdot 10^{-3}$ & $ 1.3 \cdot 10^{-3}$ \\
2.0 kbps & $1.2 \cdot 10^{-3}$ & $3.5 \cdot 10^{-3}$ \\
2.4 kbps & $ 1.3 \cdot 10^{-2}$ & $ 2.6 \cdot 10^{-2}$ \\
2.8 kbps & $ 3.4 \cdot 10^{-2}$ & $6.6 \cdot 10^{-2}$ \\
3.2 kbps & $ 1.0 \cdot 10^{-1}$ & $ 1.3 \cdot 10^{-1}$ \\
3.6 kbps & $1.2 \cdot 10^{-1}$ & $1.9 \cdot 10^{-1}$ \\
4.0 kbps & $ 2.0 \cdot 10^{-1}$ & $ 2.6 \cdot 10^{-1}$ \\ \hline
\end{tabular}
}
\label{ser_3G}
\end{table}

\begin{table}[h!]
\centering
\caption{Symbol error rate of DoV signal over VoIP. }
\begin{tabular}{|l|c c c c|}
\hline
\multicolumn{5}{|c|}{8 harmonics, symbol duration 2.5 ms, 4 s training period} \\ \hline
bitrate & Face Time & Skype & Signal Messenger & WhatsApp  \\ \hline
4.0 kbps &   $< 1.0 \cdot 10^{-4}$ & $< 1.0 \cdot 10^{-4}$ & $1.0 \cdot 10^{-4}$  & $ 9.6 \cdot 10^{-4}$    \\ 
4.8 kbps &   $< 1.0 \cdot 10^{-4}$ & $1.0 \cdot 10^{-4}$ & $9.3 \cdot  10^{-4}$  & $5.1 \cdot  10^{-3}$   \\
5.6 kbps &  $< 1.0 \cdot 10^{-4}$ & $1.2 \cdot 10^{-4}$  & $4.4 \cdot 10^{-3}$ & $1.7 \cdot 10^{-2}$    \\  
6.4 kbps &  $6.7 \cdot 10^{-4}$ & $3.0 \cdot 10^{-3}$ & $6.2 \cdot 10^{-2}$ & $8.6 \cdot 10^{-2}$    \\ \hline
\end{tabular}
\label{ser_VoIP}
\end{table}

\begin{figure}[h]
\centering
 \subfloat{
 {\includegraphics[width=1\textwidth]{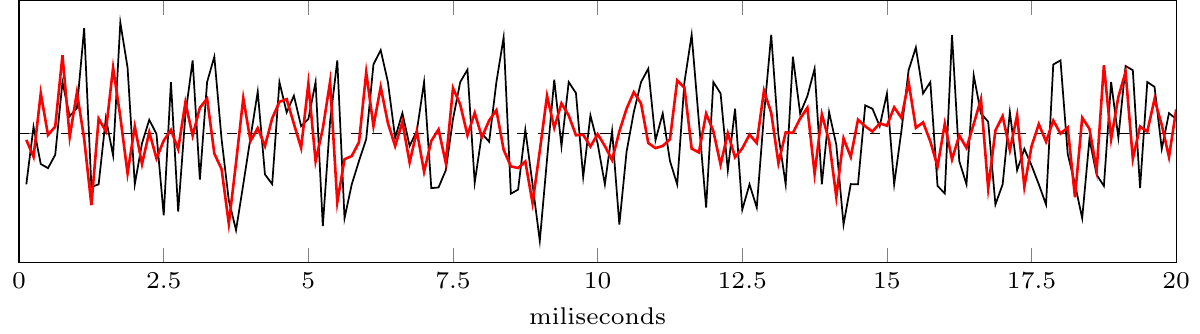}}
 }
  \caption{DoV signal at the bitrate 2.8 kbps, before (black line) and after (red line) transmission over the 3G network. The fragment displays eight consecutive DoV symbols of duration 2.5 ms consisting of 8 harmonics at frequencies 400~Hz, 800 Hz, ..., 3200 Hz, with a 400 Hz step.}
  \label{DoV_rec}
\end{figure}

In the case of 3G connection, the overall symbol error rates given in Table \ref{ser_3G} are higher compared to the simulation results presented in Fig. \ref{SER}. Additional signal distortion is possibly caused by several signal processing stages in the phones and also by multiple voice compression in the network \citep{katugampala2003secure}. Nevertheless, the DoV signal based on faster modulation and smaller codebook sizes again demonstrated lower error rates. Finally, the results emphasize the importance of voice channel estimation, which significantly improves the symbol error rate. Figure~\ref{DoV_rec} displays the small fragment of the DoV signal sent over the 3G channel.

Contrary to 3G, VoIP enables very high DoV bitrates, up to full OFDM narrowband transmission at 6.4 kbps. The improved results provided in Table \ref{ser_VoIP} are achieved due to mild signal distortion given by high throughput and network stability. However, since VoIP is a packet-based system without any guarantee of Quality of Service (QoS), short interruptions in the network connection may cause many packet dropouts. The negative impact of dropouts is typically mitigated by the re-synthesis of lost frames by VoIP application, leading to non-recoverable damages to the DoV signal and hindering the system's re-synchronization. 

\section{Secure Voice Communication}
\label{Section_voice}

This section provides a detailed proposition of a scheme for secure voice communication over 3G and VoIP, using small portable devices with limited battery capacity. The system has been successfully tested in a controlled, real-world environment and with pre-computed DoV signals. The performance results are followed by a short discussion on security and computational complexity.

\subsection{Communication system}

\begin{figure}[h]
\centering
    \includegraphics[width=0.9\textwidth]{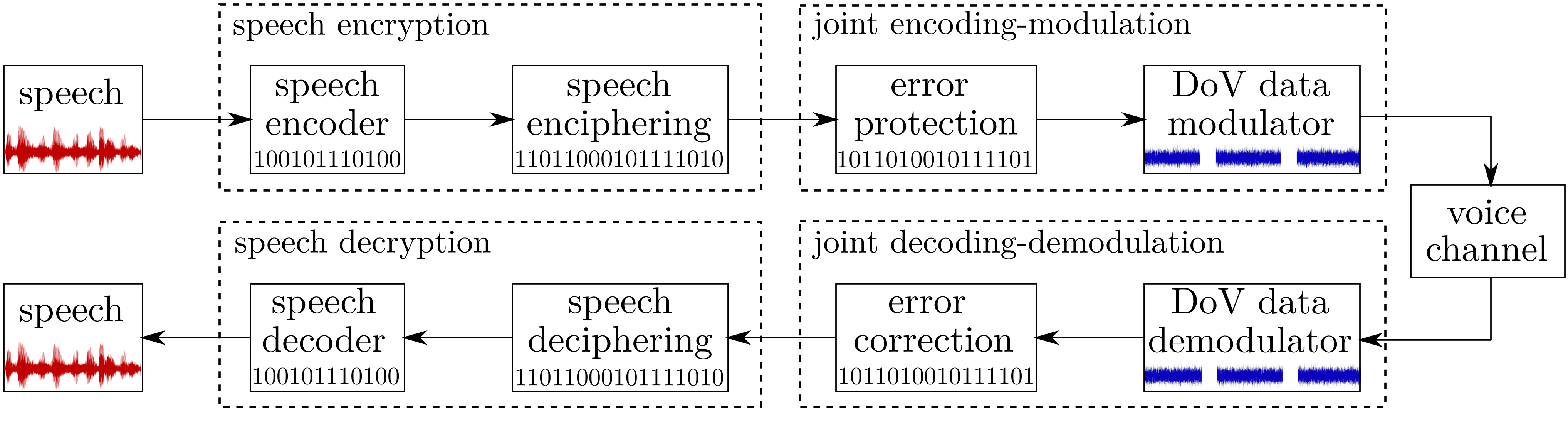}
    \caption{Encrypted speech over voice channel scheme.}
    \label{dov}
\end{figure}

Figure \ref{dov} presents a simplified diagram of a system for secure voice communication over a voice channel, which transforms consecutive portions of speech into DoV frames of the same duration. The scheme substantially resembles a classical digital communication system: it consists of speech encoding, followed by encryption, error correction, and data modulation blocks. Although the input and output signals of the processing chain are analog, all internal processing is performed digitally.

The system settings should be a trade-off between operational constraints (restricted bandwidth, real-time processing, synchronization) and the desired security level against eavesdroppers and active attackers from within the network. Depending on the voice channel type, two modes of operation may be considered: a~low mode designed for 3G cellular calls and a high mode for VoIP. The system parameters selected in the following experiments are presented in Table \ref{system_parameters} and are used only for illustration.

\begin{table}[h]
\centering
\caption{Selected parameters of the secure voice communication system.}
\begin{tabular}{|l|c|c|}
\hline
version: & low mode ( 3G )& high mode ( VoIP )\\ \hline \hline
\textbf{DoV frame} & & \\
codebook size: & 64 & 4096 \\
DoV symbol order: & 6 bits & 12 bits \\ 
modulation rate: & 400 baud & 400 baud \\
bitrate: & 2400 bps & 4800 bps \\  
frame duration: & 80 ms & 60 ms \\
frame length: & 32 symbols / 192 bits & 24 symbols / 288 bits \\
\textbf{Reed-Solomon coding} & & \\
RS symbol order: & 6 bits & 6 bits \\
message length: & 20 symbols / 120 bits & 28 symbols / 168 bits \\
 $\cdot$ encrypted speech: & 96 bits & 144 bits \\
 $\cdot$ frame counter: & 16 bits & 16 bits \\
 $\cdot$ control checksum: & 8 bits & 8 bits \\
code length: & 28 symbols / 168 bits & 40 symbols / 240 bits \\ 
redundancy: & 8 symbols / 48 bits & 12 symbols / 72 bits \\
\textbf{Voice enciphering}  & AES 256 (CTR mode) & AES 256 (CTR mode) \\ 
\textbf{Voice compression}  & Codec2 1200 bps & Codec2 2400 bps \\ \hline
\end{tabular}
\label{system_parameters}
\end{table}

The processing chain starts with low-bitrate speech compression. In this work, voice is encoded by Codec2, an open-source algorithm developed by Rowe\footnote{\url{https://rowetel.com}} and J.-M. Valin, which offers speech compression down to 450 bps \citep{erhardt2019open}. In the next step, the encoded voice frames are enciphered by AES in the counter mode of operation and with a secret key of 256 bits with a random initial value (IV). 

The encrypted binary stream is protected against channel errors by shortened Reed-Solomon (RS) codes with erasures \citep{lin2001error,neubauer2007coding} and 6-bit symbols. The error correction capabilities of RS codes depend only on the redundancy length, which is not the case for Turbo and LDPC codes \citep{tahir2017ber}. Moreover, non-binary symbol processing of RS codewords seems suitable for symbol-to-symbol demodulation of the DoV signal. In particular, one or more RS symbols can be represented by a single DoV symbol.

Erasure decoding improves correction capabilities of RS codes, provided that the localization of errors are known. The demodulator may try to guess the erroneous symbols, using a straightforward metric that considers symbol energy and its distance to the closest codebook symbol. Thus, when the first decoding attempt fails, the decoder may reiterate decoding with new estimated erasure positions until the 8-bit control checksum (8-CRC) matches.

\begin{figure}[h]
\centering
\subfloat{
    \includegraphics[width=0.9\textwidth]{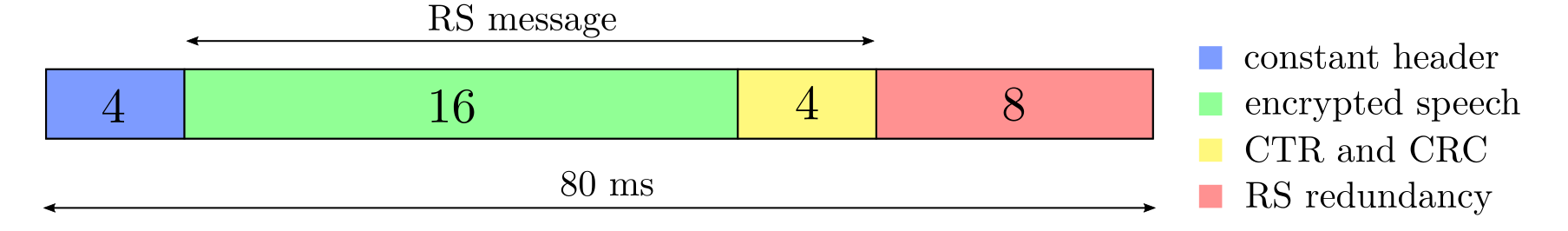}
}
\\
\subfloat{
    \includegraphics[width=0.9\textwidth]{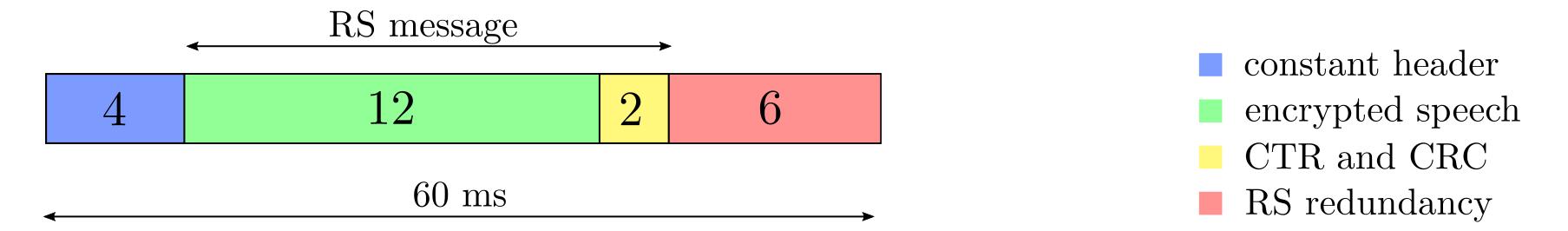}
}
    \caption{DoV frame structure in a low (up) and a high (bottom) mode of operation. The numbers indicate the lengths of frame sections, given as a cardinality of DoV symbols. In the high mode, one DoV symbol represents two RS symbols. }
    \label{RS_blocks}
\end{figure}

In the proposed scheme, each RS codeword is directly encoded into one DoV frame, as described in Fig. \ref{RS_blocks}. A constant header and a counter (CTR) enable decoding and decryption of DoV frames independently from each other, simplifying the re-synchronization in the presence of signal dropouts. Extensive experiments have shown that a 10-ms header is usually sufficiently long to keep signal synchronization or detect a DoV frame after signal restoration. In addition, the 16-bit counter permits re-synchronization after more than one hour of lost connection. 

The duration of a DoV frame is equal to the portion of speech encoded by this frame, which is a valid requirement for real-time communication. Selected voice compression rates, 1.2 kbps, and 2.4 kbps depending on the mode, are low enough to append error correction redundancy at the end of each DoV frame. 

The system was tested over cellular and VoIP calls. Table \ref{ser_voice} presents the decoding results of several minutes of speech recording sent through using the 4G mobile data connectivity between two iPhones 6 registered to different network operators. The effective bit error rates (BER) and frame error rates (FER) take into account errors due to system de-synchronizations and short signal dropouts.

\begin{table}[h]
\centering
\caption{Performance of encrypted voice transmission over cellular voice channels and VoIP.}
\begin{tabular}{|l|c c c c c|}
\hline
& 3G & Face Time & Skype & Signal Messenger & WhatsApp  \\ \hline \hline
effective BER: & $3.7 \cdot 10^{-3}$ & $< 1.0 \cdot 10^{-4}$  & $< 1.0 \cdot 10^{-4}$ & $< 1.0 \cdot 10^{-4}$  & $7.8 \cdot 10^{-4}$  \\ 
effective FER: & $1.9 \cdot 10^{-2}$ &  $< 1.0 \cdot 10^{-3}$ & $< 1.0 \cdot 10^{-3}$ & $< 1.0 \cdot 10^{-3}$ & $2.1 \cdot 10^{-3}$ \\ \hline
\end{tabular}
\label{ser_voice}
\end{table}

\begin{figure}[h!]
\centering
  \subfloat{
 {\includegraphics[width=1\textwidth]{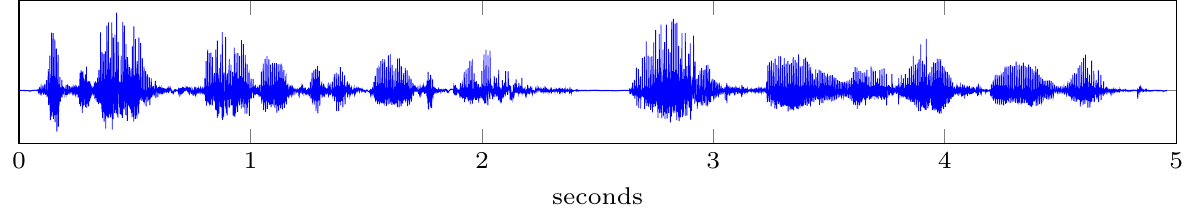}}
\label{step1}
  }
  \\
 \subfloat{
{\includegraphics[width=1\textwidth]{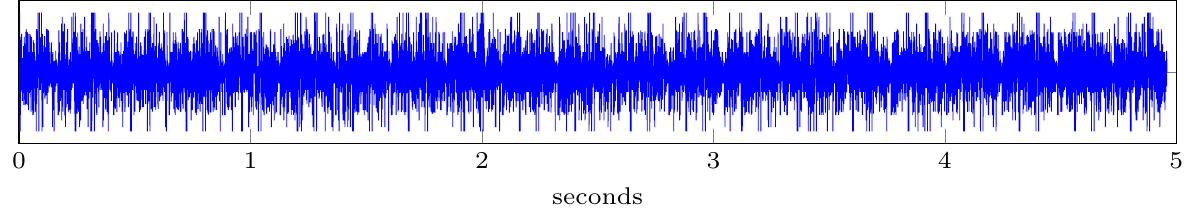}}
\label{step2}
 }
 \\
 \subfloat{
{\includegraphics[width=1\textwidth]{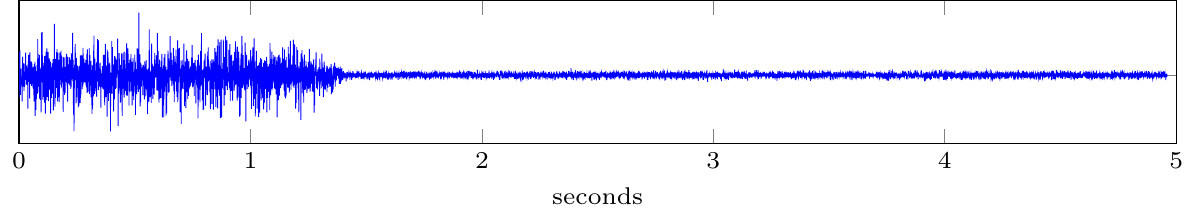}}
\label{step3}
 }
 \\
 \subfloat{
{\includegraphics[width=1\textwidth]{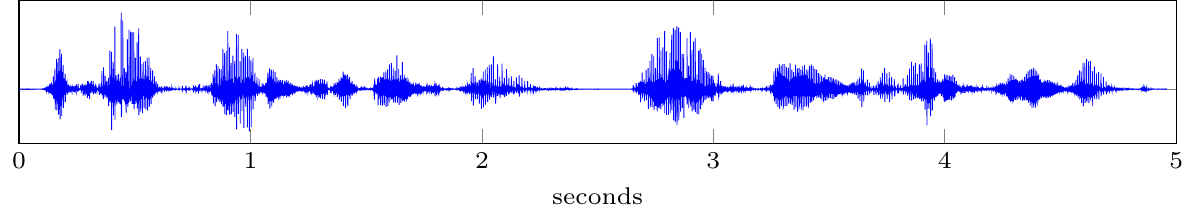}}
\label{step4}
 }
  \caption{Consecutive stages of the signal in secure voice communication over a 3G call. From top to bottom: the initial speech, the sent DoV signal, the received DoV signal and the re-synthesized speech. The received signal was fully decodable despite strong signal attenuation.}
  \label{3G_voice}
\end{figure}

Figure \ref{3G_voice} shows the consecutive waveforms of a signal processed by a 3G network. The initial speech waveform presented in Fig. \ref{step1} is compressed, encrypted, and encoded into the DoV signal of equal duration in Fig. \ref{step2}. The received signal displayed in Fig. \ref{step3} is strongly attenuated after less than 2 seconds of transmission, classified by the Voice Activity Detector (VAD) as non-speech-like. However, correct decoding is still possible as long as the harmonic structure of the signal is preserved, as shown in Fig. \ref{step4}.

\begin{figure}[h!]
\centering
 \subfloat{
{\includegraphics[width=1\textwidth]{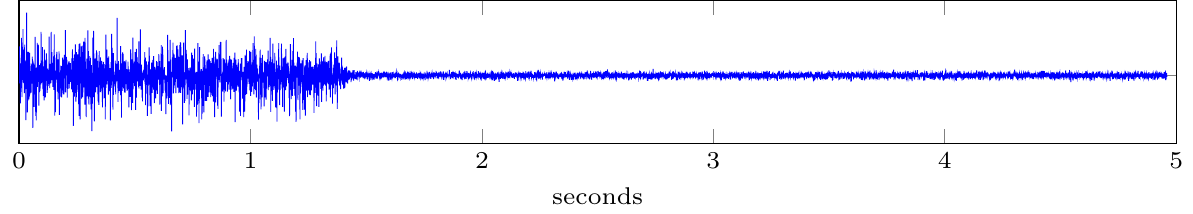}}
 }
 \\
 \subfloat{
{\includegraphics[width=1\textwidth]{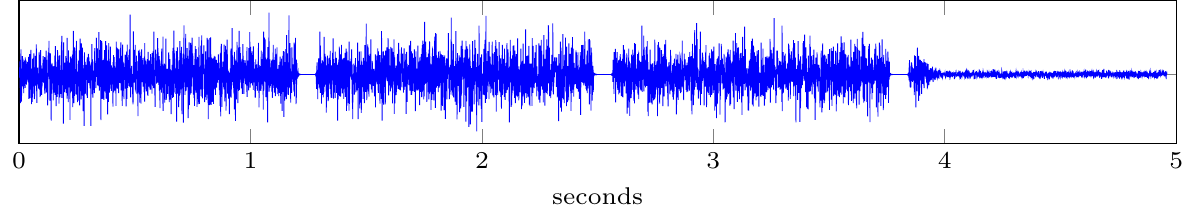}}
 }
  \caption{Comparison of the received DoV signal in a 3G call (top) without and (bottom) with silence insertion every 16th frame. Depending on the connection type and the silence insertion rate, this technique may postpone or prevent signal suppression.}
  \label{3G_voice_silence}
\end{figure}

The distortion of the received signal can vary, depending on the network type and the phones used for communication. To counteract the blockage of stationary signals by VAD and Noise Suppression, several authors suggest to alternate two DoV codebooks defined over two non-overlapping bandwidths \citep{shahbazi2010data,sapozhnykov2012low}. This work proposes another complementary technique: periodic silence insertion in place of some DoV frames, as depicted in Fig. \ref{3G_voice_silence}. It was observed that depending on the chosen rate of silence insertion and the type of connection, these silences significantly postpone or even prevent signal suppression. On the reception side, these inserted silences can be classified as lost frames and re-synthesized.

\subsection{Security discussion}

The presented voice communication scheme must offer sufficiently high levels of secrecy and authentication
in order to prevent speech interception. A major risk is the recording and off-line cryptanalysis of the network traffic by passive eavesdroppers. Securing the communication against eavesdroppers is especially important because the encrypted and non-speech signal can be easily detected by some advanced Data Leakage Prevention (DLP) and Content Monitoring and Filtering (CMF) systems protecting against unauthorized data extrusion \citep{chae2016privacy,hauer2015data,lee2017vulnerability}. Active attackers controlling the network are more likely to block or distort the fragile DoV signal, which is technically very simple. However, a powerful and knowledgeable attacker who can synthesize a compatible DoV signal in real-time may modify the signal or insert its own.    

The chosen AES cipher in the counter mode of operation, if implemented correctly, is believed to provide security against passive eavesdroppers \citep{lipmaa2000ctr,jonsson2002security}. On the other hand, enciphering in counter mode does not guarantee data integrity \citep{katz2014introduction}, giving some space for adversarial manipulations. Therefore, the common practice is to combine the AES in counter mode with a cryptographic message authentication function \citep{housley2004rfc3686}. Unfortunately, due to severe bandwidth limitations appending the authentication check is not viable. Instead, it would be possible to randomly shuffle the positions of encrypted bits within one DoV frame \citep{morris2009encipher,stefanov2012fastprp}. The motivation for this is to prevent malicious attackers from intentional modifications of the transmitted content. While still capable of replacing several DoV symbols, the attacker should not benefit from distorting the transmitted signals.

Finally, it is assumed that both users share a common secret cryptographic key used for encryption. Secure key exchange can become challenging when the voice channel is the only available communication channel. With decentralized implementations of the proposed system, there would be no practical possibility to add the Trusted Third Party for user's authentication. A few protocols overcome this limitation by using vocal verification \citep{pasini2006sas,callas2011zrtp,krasnowski2020introducing}. In such a scenario, users compare freshly generated random strings vocally while challenging another speaker's voice profile. 

\subsection{Computational complexity}

The goal of real-time operation on small portable devices puts a big emphasis on computational optimization of the proposed system. It can be noticed that PSK-OFDM modulation \citep{3gpp2010lte}, AES-CTR encryption \citep{housley2004rfc3686,park2011efficient}, Reed-Solomon error correction \citep{biard2008reed,3gpp2003universal} and the speech encoding \citep{wisayataksin2019efficient} algorithms mentioned in this work have been already widely adopted in wireless communication with mobile phones or in the computationally constrained environment, including real-time applications. However, the presented system was implemented in GNU Octave environment,\footnote{\url{https://www.gnu.org/software/octave/}} serving as a proof-of-concept only. There is still considerable work to be done to efficiently integrate all these elements into a single system operating on a device with limited resources, like mid-range smartphones.

\section{Conclusion}
\label{Section_conclusion}

In this article, we detailed a new and versatile Data over Voice technique for secure voice communications over LPC-based voice channels, like cellular networks and VoIP. Based on codebooks with harmonic symbols, the proposed solution is well-grounded on the fundamental principles of LPC coding.

A thorough analysis of OFDM signals compressed by some prominent voice coders revealed that the distortion is statistically close to a symmetric bivariate Gaussian distribution over the complex phase-amplitude plane. However, this distortion is not uniformly distributed in the spectral domain. Thus, we proposed an optimized demodulation metric based on spectrally weighted Euclidean distance with phase shift correction.

The tedious design process of DoV codebooks has been considerably simplified by using quaternary error correction codes. With OFDM symbols being treated as codes over a quaternary ring, codebook construction reduces to finding a set of quaternary codes that maximizes the minimum Lee distance.

The performance of our DoV technique has been evaluated through simulations and real-world tests over real voice connections between two mobile phones. A bitrate of 2.4 kbps over 3G call and 6.4 kbps over VoIP have been achieved with acceptably low symbol error rates. These tests highlight the need to properly characterize the channel distortion before transmission properly.

Finally, the work described a scheme for secure voice communications over voice channels in high and low bitrate modes of operation. The system has been practically validated for real-time voice transmission over cellular networks and VoIP with small effective bit error rates. To mitigate the negative impact of VAD, we also proposed a new method based on the insertion of repetitive silences.

The promising results presented in this work suggest some further investigation of the proposed DoV technique. A big emphasis has to be put on signal synchronization on the reception side and reducing the computational cost of signal demodulation. Additionally, sensible codebook structuring, combined with the exploitation of phase symmetries, may significantly lower the number of correlations in a demodulator.

\section{Acknowledgments}

This work is supported by grant DGA Cifre-Defense program No 01D17022178 DGA/DS/MRIS and AID program No~SED0456JE75.

\bibliography{bibliography} 

\end{document}